\documentclass{article}
\usepackage{graphicx}%style pour inserer figures en eps
\usepackage{here}%pour inserer figures et tables a emplacement
\def\beq{\begin{equation}}
\def\eeq{\end{equation}}
\def\bea{\begin{eqnarray}}
\def\eea{\end{eqnarray}}
\def\bq{\begin{quote}}
\def\eq{\end{quote}}

\def\bear{\begin{array}}
\def\ear{\end{array}}
\def\nnb{\nonumber}
\def\ga{\left(}
\def\dr{\right)}
\def\aga{\left\{}
\def\adr{\right\}}

\def\rar{\rightarrow}
\def\Lrar{\Longrightarrow}

\def\nnb{\nonumber}
\def\la{\langle}
\def\ra{\rangle}
\def\nin{
\vspace*{0.5cm}

\noindent}
\def\ba{\begin{array}}
\def\ea{\end{array}}
\def\bm{\overline{m}}

\def\als{\alpha_s}
\def\as{\ga \frac{\alpha_s}{\pi}\dr}
\def\asb{\ga \frac{\overline\alpha_s}{\pi}\dr}
\def\alsb{\overline{\alpha}_s}
\def\mb{\overline{m}}
\def\msb{\overline{MS}}

\def\pr{\partial}

\def\gam5{\gamma_5}

\begin{document}
\pagestyle{empty}
\begin{flushright}
PM 02/06
\end{flushright}
\vspace*{1cm}
%\begin{center}
%%%%%%%%%%%%%%%%%%%%%%%%%%%%%%%%%%
\subsection*{LIGHT AND HEAVY QUARK MASSES,\\
FLAVOUR BREAKING OF CHIRAL CONDENSATES,
\\ MESON WEAK LEPTONIC
   DECAY CONSTANTS IN QCD
\footnote{This review updates and completes the reviews
\cite{SNL,SNSRH} and some parts of the book \cite{SNB}. It
has been extracted from a chapter of the forthcoming book:
{\it QCD as a theory of hadrons: from partons to confinement} \cite{SNB02}.}}.
%%%%%%%%%%%%%%%%%%%%%%%%%%%%%%%%%%
\vspace*{1.cm}
\nin
{\bf Stephan Narison} \\
%\vspace{0.3cm}
%\nin
Laboratoire de Physique Math\'ematique,
Universit\'e de Montpellier II,\\
Place Eug\`ene Bataillon,
34095 - Montpellier Cedex 05, France\\
Email: qcd@lpm.univ-montp2.fr
\vspace*{1.cm}
\nin
{\bf Abstract} \\
%\end{center}
%\vspace*{2mm}
\noindent
We review the present status for the determinations of the light and
heavy quark masses, the
light quark chiral condensate and the decay constants of light and heavy-light
(pseudo)scalar mesons from QCD spectral sum rules (QSSR). Bounds
on the light quark running masses at 2 GeV are found to be (see
Tables \ref{tab: lowbound} and \ref{tab: upbound}): 6
MeV$<(\overline{m}_d+\bm_u)(2)<$11~MeV and 71 MeV$<\bm_s(2)<$148 MeV. The
agreement of the ratio $m_s/(m_u+m_d)=24.2$ in Eq. (\ref{eq: ms/mud})
from pseudoscalar sum rules with the one $(24.4\pm
1.5)$ from ChPT indicates the consistency of the pseudoscalar sum 
rule approach.
QSSR predictions from different channels for the light
quark running masses lead to (see Section \ref{sec: lsum}):
$\overline{m}_s(2)=(117.4\pm 23.4)$~MeV,
$(\overline{m}_d+\bm_u)(2)=(10.1\pm 1.8)$~MeV,
$(\bm_d-\overline{m}_u)(2)=(2.8\pm 0.6)$~MeV with the corresponding
values of the RG invariant masses.
The different QSSR predictions for the heavy quark masses lead to the
running
masss values: $\bm_c(\bm_c)=(1.23\pm 0.05)$ GeV and
$\bm_b(\bm_b)=(4.24\pm 0.06)$~GeV (see Tables
\ref{tab: mc} and \ref{tab: mb}), from which one can extract the
scale independent ratio $m_b/m_s=48.8\pm
9.8$. Runned until $M_Z$, the $b$-quark mass becomes:
$\bm_b(M_Z)=(2.83\pm 0.04)$~GeV in good agreement with the average of
direct measurements
$(2.82\pm 0.63)$~GeV from three-jet heavy quark production
at LEP, and then supports the QCD running predictions based on the
renormalization group equation.
As a result, we have updated our old predictions of the weak decay constants
$f_{\pi'(1.3)}, ~f_{K'(1.46)}$,
  $f_{a_0(0.98)}$ and $f_{K^*_0(1.43)}$ (see Eqs. (\ref{eq: fpseu})
and (\ref{eq: fscal})). We obtain from a global fit of the light 
(pseudo)scalar  and $B_s$
mesons, the flavour breakings of the {\it normal ordered} chiral 
condensate ratio:
$\la\bar ss\ra/\la\bar uu\ra=0.66\pm 0.10$ (see Eq. (\ref{eq: 
ssdd})). The last section is
dedicated to the QSSR determinations of
$f_{D_{(s)}}$ and
$f_{B_{(s)}}$.

\vspace*{1.0cm}
\begin{flushleft}
PM 02/06\\
%\today
February 2002
\end{flushleft}
\vfill\eject
\setcounter{page}{1}
   \pagestyle{plain}

%%%%%%%%%%%%%%%%%%%%%%%%%%%%%%%%%%
\section{Introduction}
One of the most important parameters of the standard model and chiral
symmetry is the light and heavy
quark masses. Light quark masses and chiral condensates are useful
for a much better understanding of the
realizations of chiral symmetry breaking
\cite{WEINBERG}--\cite{14} and for some eventual explanation of the origin of
quark masses in unified
models of interactions \cite{FRITZ}. Within some popular
parametrizations of the hadronic matrix
elements \cite{BURAS}, the strange quark mass can also largely
influence the Standard Model prediction of
   the
$CP$ violating parameters
$\epsilon'/\epsilon$, which have been mesured recently
\cite{NA48}.
However, contrary to the QED case where leptons are observed, and
then the physical masses
can be identified with the pole of the propagator (on-shell mass
value)\footnote{For a first explicit definition of the
perturbative quark pole mass in the $\overline{MS}$-scheme, see
\cite{COQUE,TARRA1} (renormalization-scheme
invariance) and \cite{SN1} (regularization-scheme invariance).}, the
quark masses are difficult to define
because of confinement which does not allow to observe free quarks.
However, despite this difficulty, one can consistenly
treat the quark masses in perturbation theory
like the QCD coupling constant.They obey a differential equation, where
its boundary condition can be identified with the renormalized mass
of the QCD lagrangian. The corresponding solution is
the running mass, which is gauge invariant but renormalization scheme
and scale dependent, and the associated renrmalization group
invariant mass. To our knowledge, these
notions have  been introduced for the first time in
\cite{FLORATOS}. In practice, these masses are conveniently defined
within the standard
$\overline{MS}$-scheme discussed in previous chapters. In addition to
the determination of the ratios of light quark
masses (which are scale independent) from current algebra
\cite{WEINBERG}, and from chiral perturbation theory (ChPT), its
modern version \cite{WEINChPT}--\cite{ChPT},
a lot of effort reflected in the literature
\cite{PDG} has been put into extracting directly from the data the
running quark masses
using the SVZ \cite{SVZ} QCD spectral sum rules (QSSR) \cite{SNB},
LEP experiments and lattice simulations.
The content of these notes is:
\begin{itemize}
\item a review of the light and heavy quark mass
determinations from the different QCD
approaches.
\item a review of the direct determinations of the quark
vacuum condensate using QSSR and an update of the analysis
of its flavour breakings using a global fit of the meson systems.
\item An update of the determinations of
the light (pseudo)scalar decay constants, which, in particular, are useful
for understanding the $\bar qq$ contents of the light scalar mesons.
\item A review of the determinations
of the weak leptonic decay of the heavy-light pseudoscalar mesons 
$D_{(s)}$ and $B_{(s)}$.
\end{itemize}
This review develops and updates
the review papers \cite{SNL,SNSRH} and some parts of the book 
\cite{SNB}. It also updates
previous results from original works.
%%%%%%%%%%%%%%%%%%%%%%%%%%%%%%%%%%%%%%%%%%%%%%%%%%%%%%%%%%%%%%%%%%%%%%%
\section{Definitions of perturbative quark masses in QCD}
Let's remind the meaning of quark masses in QCD.
One starts from the mass term of the QCD lagrangian:
\beq
{\cal L}_m=m_i\bar\psi_i\psi_i~,
\eeq
where $m_i$ and $\psi_i$ are respectively the quark mass and field.
It is convenient to introduce the {\it dimensionless mass}
$x_i(\nu)\equiv m_i(\nu)/\nu$, where
$\nu$ is the renormalization scheme subtraction constant. The running
quark mass is a solution
of the differential equation:
\beq
\frac{d\overline{x}_i}{dt}=(1+\gamma(\alpha_s))\overline{x}_i(t)~:\overline{x}_i(t=0)=x_i(\nu)~.
\eeq
In the $\overline{MS}$-scheme, its solution to order $a_s^3$
($a_s\equiv \alpha_s/\pi$) is:
\bea
\bm_i(\nu)&=&\hat{m}_i\ga -\beta_1 a_s(\nu)\dr^{-\gamma_1/\beta_1}
\Bigg\{1+\frac{\beta_2}{\beta_1}\ga \frac{\gamma_1}{\beta_1}-
   \frac{\gamma_2}{\beta_2}\dr a_s(\nu)~
+\frac{1}{2}\Bigg{[}\frac{\beta_2^2}{\beta_1^2}\ga \frac{\gamma_1}
{\beta_1}-
   \frac{\gamma_2}{\beta_2}\dr^2-
\frac{\beta_2^2}{\beta_1^2}\ga \frac{\gamma_1}{\beta_1}-
   \frac{\gamma_2}{\beta_2}\dr\nnb\\&+&
\frac{\beta_3}{\beta_1}\ga \frac{\gamma_1}{\beta_1}-
   \frac{\gamma_3}{\beta_3}\dr\Bigg{]} a^2_s(\nu)+
1.95168a_s^3\Bigg\}~,
\eea
where $\gamma_i$ and $\beta_i$ are the ${\cal{O}}(a_s^i)$ coefficients of the
quark-mass anomalous dimension and $\beta$-function, which read
for $SU(3)_c\times SU(n)_f$:
\bea
\gamma_1&=&2~,~~~\gamma_2=\frac{1}{6}\ga {101\over 2}-{5n\over
2}\dr~,~~~\gamma_3={1\over 96}\Big{[}3747-\ga 160\zeta_3-\frac{2216}{ 9}\dr n-
{140\over 27} n^2\Big{]}~\nnb\\
\beta_1&=&-{1\over 2}\ga 11-{2\over 3}n\dr~,~~~\beta_2=-{1\over 4}\ga
51-{19\over 3}n\dr~,~~~\beta_3=-{1\over 64}\ga 2857-{5033\over
9}n+{325\over 27}n^2\dr~.
\eea
The invariant mass $\hat{m}_i$ has been introduced for the first time
by \cite{FLORATOS} in connection
with the analysis of the breaking of the Weinberg sum rules by the
quark mass terms in QCD.
For the heavy quarks, one can also define a perturbative (short
distance) pole mass at the pole
of the propagator. The IR finiteness of the result to order $\als^2$
has been explicitly
shown in \cite{COQUE,TARRA1}. The independence of $M_{pole}$
on the choice of the regularization-scheme has been demonstrated
in \cite{SN1}.
The extension of the previous result to order $\als^2$ is: \cite{GRAY}:
\bea
M_{pole}&=&\mb(p^2)\Bigg{[}1+\ga\frac{4}{3}+\ln{\frac{p^2}{m^2}}\dr\as+\nnb\\
&&\Bigg{[}K_Q+\ga \frac{221}{24}-\frac{13}{36}n\dr
\ln{\frac{p^2}{m^2}}+\ga\frac{15}{8}-\frac{n}{12}\dr
\ln^2{\frac{p^2}{m^2}}\Bigg{]}\as^2\Bigg{]}~,
\eea
%\end{document}
where in the RHS $m$ is the running mass evaluated at $p^2$ and:
\beq
K_Q=17.1514-1.04137n+\frac{4}{3}\sum_{i\not=Q}\Delta\ga r\equiv
\frac{m_i}{M_Q}\dr.
\eeq
For $0\leq r\leq 1$, $\Delta(r)$ can be approximated, within an
accuracy of 1$\%$
by:
\beq
\Delta(r) \simeq \frac{\pi^2}{8}r-0.597r^2+0.230r^3~.
\eeq
It has been argued that the pole masses can be affected by
nonperturbative terms induced by the resummation of the QCD
perturbative series \cite{BENEKE} and alternative definitions free
from such ambiguities
have been proposed (residual mass \cite{POLE3} and 1S mass
\cite{HOANG}). Assuming that the QCD potential has no
linear power corrections, the residual
or potential-subtracted (PS) mass is related to the pole mass as:
\beq
M_{PS}=M_{pole}+{1\over 2}\int_{\vert\vec q\vert<\mu}\frac{d^3\vec
q}{(2\pi)^3}V(\vec q)~.
\eeq
The 1S mass is defined as half of the perturbative component to the
$^3S_1$ $\bar QQ$ ground state, which is half
of its static energy $ \la 2M_{pole}+V \ra$ \footnote{These definitions
might still be affected by a dimension--2 term advocated in 
\cite{ZAKA,CNZ,SNZAK}, which might
limit their accuracy \cite{ZAKAP}.}.
The running and short distance pole mass defined at a given order of
PT series will be used in the following
discussions.
%%%%%%%%%%%%%%%%%%%%%%%%%%%%%%%%%%
\section{Ratios of light quark masses from ChPT}
%\begin{figure}[hbt]
%\begin{center}
%\includegraphics[width=7cm]{chpt.ps}
%\caption{$m_s/m_d$ versus $m_u/m_d$ from \cite{14}. }
%\label{fig:largenenough}
%\end{center}
%\end{figure}
%\nin
The ratios of light quark masses are well-determined from current
algebra \cite{WEINBERG}, and ChPT
\cite{WEINChPT}. In this approach, the meson masses are expressed
using a systematic expansion in terms of
the light quark masses:
\bea
   M^2_{\pi^+}&=& (m_u+m_d){ B}  +{\cal O}(m^2)+...\nnb\\
    M^2_{K^+}&=& (m_u+m_s){ B}  +{\cal O}(m^2)+...\nnb\\
    M^2_{K^0}&=& (m_d+m_s){ B}  +{\cal O}(m^2)+...
\eea
where {$ B\equiv -\la\bar\psi\psi\ra/f^2_K$ from the Gell-Mann, Oakes, Renner
relation \cite{GMOR}}:
\beq
m^2_\pi f^2_\pi\simeq -(m_u+m_d)\la\bar\psi\psi\ra +{\cal O}(m^2)~.
\eeq
   However, only the ratio, which is scale independent can be   well
determined. To leading   order in $m$ \cite{WEINChPT}\footnote{In
Generalized ChPT, the contribution of the
$m^2$-term can be as large as the $m$ one \cite{GCHPT}, which
modifies drastically these ratios.}:
\bea
\frac{m_u}{m_d}& \approx&\frac{M^2_{\pi^+}-M^2_{K^0}+M^2_{K^+}}
{M^2_{\pi^+}+M^2_{K^0}-M^2_{K^+}}\approx  0.66\nnb\\
\frac{m_s}{m_d}&\approx&\frac{-M^2_{\pi^+}+M^2_{K^0}+M^2_{K^+}}
{M^2_{\pi^+}+M^2_{K^0}-M^2_{K^+}}\approx  20
\eea
Including the next order + electromagnetic corrections, the ratios of
masses are constrained on the ellipse:
\beq
\ga\frac{m_u}{m_d}\dr^2+\frac{1}{Q^2}\ga\frac{m_s}{m_d}\dr^2=1
   \eeq
where: $Q^2\simeq (m^2_s-\hat{m}^2)/(m^2_d-m^2_u)=22.7\pm 0.8$ using
the value of the
$\eta\rar\pi^+\pi^-\pi^0$ from the PDG average \cite{PDG}, though
this value can well be in the range
22--26, to be compared with the Dashen's formula \cite{DASHEN} of
24.2; $\hat{m}\equiv (1/2)(m_u+m_d)$.
In the figure of \cite{14}\footnote{This figure will be shown in the 
book version of the paper.},
one shows the range spanned by $R\equiv (m_s-\hat{m})/(m_d-m_u)$ and
the corrections to the GMO
mass formula $\Delta_M:~M^2_8=(1/3)(4M^2_K-m^2_\pi)(1+\Delta_M)$. The
Weinberg mass ratio \cite{WEINBERG} is
also shown which corresponds to the Dashen's formula  and $R\simeq
43$. At the intersection of
different ranges, one deduces \cite{14}:
\label{chpt}
\bea
&&\frac{m_u}{m_d}= 0.553\pm 0.043~,~\frac{m_s}{m_d}=
18.9\pm 0.8,\nnb\\
&&\frac{2m_s}{(m_d+m_u)}= 24.4\pm 1.5.
\eea
The possibility to have a $m_u=0$ advocated in \cite{MANO} appears to
be unlikely as it
implies too strong flavour symmetry breaking and is not supported by
the QSSR results
from 2-point correlators of the divergences of the axial and vector
currents, as will be shown in the
next sections.

%%%%%%%%%%%%%%%%%%%%%%%%%%%%%%%%%%%%%%%%%
\section{Bounds on the light quark masses}
In QSSR, the estimate and lower bounds of the sum
of the light quark masses from the
pseudoscalar sum rule have been
firstly done in \cite{43,PSEUDO}, while a bound on the quark
mass difference has been firstly derived in \cite{SCAL}. The
literature in this subject
of light quark masses increases with time \footnote{Previous works are
are reviewed in \cite{SNL,SNB}.}
%See also: \cite{SNP,SNP2,CHETDPS,JAMIN},\cite{PAVERTR}--\cite{OLLER}.}.
However, it is in
some sense quite disappointing that in most of the published papers
no noticeable progress
has been done since the previous pioneering works. The most
impressive progress comes from
the QCD side of the sum rules where new calculations have become
available both on the
perturbative radiative corrections known to order $\alpha_s^3$
\cite{43,LARIN,CHETDPS} and on the nonperturbative corrections
\cite{SVZ}--\cite{JAMIN2}
\footnote{See also the chapter on two-point functions where more
references to original works are
given.}. Another new contribution is due to the inclusion of the
tachyonic gluon mass as a
manifestation of the resummation of pQCD series
\cite{ZAKA,CNZ,SNZAK}. Alas, no sharp result is
available on the exact size of direct instanton contributions
advocated to be important in
this channel \cite{IOFFE}, while \cite{GABRIELLI} claims the
opposite. Though the
instanton situation remains controversial, recent analysis
\cite{STEELE01,KAMBOR} using the results of \cite{DORGHOV} based on
the ILM of \cite{SHURYAK} indicates that this effect is
negligible justifying the neglect of this effect in different
analysis of this channel. However, it might happen that
adding together the effect of the tachyonic gluon to the one of
direct instanton might also lead to a double counting in a
sense that they can be two alternative ways for parametrizing the
nonperturbative vacuum \cite{SNZAK}. In absence of
precise control of the origin and size of these effects, we shall
consider them as new sources of errors in
the sum rule analysis.
%%%%%%%%%%%%%%%%%%%%%%%%%%%%%%%%%%%%%%%%%%%%%%%%%
\subsection{Bounds on the sum of light quark masses from pseudoscalar channels}
Lower bounds for $(\overline{m}_u+\overline{m}_{d})$ based on moments
inequalities and the
positivity of the spectral functions have been obtained, for the first time,
in \cite{43,PSEUDO}.
These bounds have been rederived recently in
\cite{LEL,YNDBOUND} to order $\alpha_s$. As cheked in \cite{SNL} for
the lowest moment and redone in \cite{KAMBOR}
for higher moments,
the inclusion of the $\alpha_s^3$ term decrease by about 10 to 15\%
the strength of these bounds,
which is within the expected accuracy of the result. \\
For definiteness, we shall discuss in details the pseudoscalar
two-point function in the $\bar us$
channel. The analysis in the $\bar ud$ channel is equivalent. It is
convenient to start from the
second derivative of the two-point function which is superficially convergent:
\begin{equation}\label{eq:disp}
\Psi''(Q^2)=\int_{0}^{\infty} dt\frac{2}{(t+Q^2)^3}\frac{1}{\pi}{\rm
Im}\Psi(t).
\end{equation}
The bounds follow from the restriction of the sum over all
possible hadronic states which can contribute to the spectral
function to the state(s) with the lowest invariant mass.
The lowest hadronic state which contributes to the
corresponding spectral function is the $K$--pole. From eq.~(\ref{eq:disp})
we then have
\begin{equation}\label{eq:psi5}
\Psi_{5}''(Q^2)= \frac{2}{(M_{K}^2 +Q^2)^3}2f_{K}^2 M_{K}^4 +
\int_{t_{0}}^{\infty} dt
\frac{2}{t+Q^2)^3}\frac{1}{\pi}\mbox{Im}\Psi_{5}(t),
\end{equation}
where $t_{0}=(M_{K}+2m_{\pi})^2$ is the threshold of the hadronic
continuum.
\\
It is convenient to introduce the moments $\Sigma_{N}(Q^2)$ of the hadronic
continuum integral
\begin{equation}\label{eq:contint}
\Sigma_{N}(Q^2)=\int_{t_{0}}^{\infty}dt\frac{2}{(t+Q^2)^3}\times
\left(\frac{t_0 +Q^2}{t+Q^2}\right)^{N}\frac{1}{\pi}{\rm Im}\Psi_{5}(t).
\end{equation}
One is then confronted with a typical moment problem (see
e.g. Ref.~\cite{AhK62}.) The positivity of the continuum spectral function
$\frac{1}{\pi}{\rm Im}\Psi_{5}(t)$ constrains the moments
$\Sigma_{N}(Q^2)$ and hence the l.h.s. of Eq. (\ref{eq:psi5}) where the light
quark masses appear. The most general constraints among the first three
moments for
$N=0,1,2$ are:
\begin{equation}
\Sigma_{0}(Q^2)\ge 0,\quad \Sigma_{1}(Q^2)\ge 0,\quad
\Sigma_{2}(Q^2)\ge 0;
\end{equation}
\begin{equation}\label{eq:diff}
\Sigma_{0}(Q^2)-\Sigma_{1}(Q^2)\ge 0,\quad
\Sigma_{1}(Q^2)-\Sigma_{2}(Q^2)\ge 0;
\end{equation}
\begin{equation}
\label{eq:quad}
\Sigma_{0}(Q^2)\Sigma_{2}(Q^2)-\left(\Sigma_{1}(Q^2)\right)^2\ge 0\,.
\end{equation}
The inequalities in Eq.~(\ref{eq:diff}) are in fact trivial
unless
$2Q^2< t_{0}$, which constrains the region in $Q^2$ to too small values
for pQCD to be applicable. The other inequalities lead however to
interesting bounds which we next discuss.
\\
The inequality $\Sigma_{0}(Q^2)\ge 0$ results in a first bound on the
running masses:
\begin{equation}\label{eq:1stbound}
\left[m_{s}(Q^2)+m_{u}(Q^2)\right]^2 \ge \frac{16\pi^2}{N_c}
\frac{2f_{K}^2 M_{K}^4}{Q^4}\times
\frac{1}{\left(1+\frac{M_{K}^2}{Q^2}\right)^3}
\frac{1}{\left[1+
\frac{11}{3}\frac{\alpha_{s}(Q^2)}{\pi} +\cdots\right]}~,
\end{equation}
where the dots represent higher order terms which have been
calculated up to
${\cal O}(\alpha_s^3)$, as well as  non--perturbative power corrections of
${\cal O}\left({1}/{Q^4}\right)$ and strange quark mass corrections of
${\cal O}\left({m_{s}^2}/{Q^2}\right)$ and
${\cal O}\left({m_{s}^4}/{Q^4}\right)$ including ${\cal O}(\alpha_s)$
terms~.   Notice that this bound is non--trivial in the large--$N_c$ limit
($f_{K}^2\sim{\cal O}(N_c)$) and in the chiral limit ($m_{s}\sim M_{K}^2$). The
bound is of course a function of the choice of the euclidean
$Q$--value at which the r.h.s. in Eq.~(\ref{eq:1stbound}) is evaluated. For
the bound to be meaningful, the choice of $Q$ has to be made sufficiently
large. In Ref.~\cite{LEL} it is shown that
$Q\geq 1.4\:{\rm GeV}$ is already a safe choice to trust the pQCD
corrections as such. The lower bound which follows from
Eq.~(\ref{eq:1stbound}) for
$m_u + m_s$ at a renormalization scale
$\mu^2=4\:{\rm GeV^2}$ results in the solid curves shown in
Fig.~\ref{fig:Fig5a}
below.
%%%%%%%%%%%%%%%%%%%
\begin{figure}[htb]
\begin{center}
\includegraphics[width=9cm]{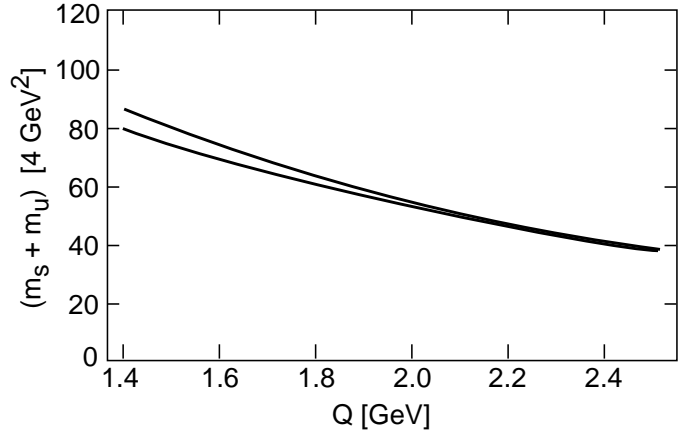}
\caption{Lower bound in {\rm MeV} to order $\als$ for
$(m_{s}+m_{u})$(2) versus $Q$ in GeV from
Eq.~(\protect\ref{eq:1stbound}) for $\Lambda_3=290$ MeV (upper curve)
and 380 MeV (lower curve). Quark mass values below the
solid curves in Fig.~\ref{fig:Fig5a} are forbidden by the bounds.}
\label{fig:Fig5a}
\end{center}
\end{figure}
%%%%%%%%%%%%%%%%%%%
\nin
The resulting value of the bound at $Q=1.4$ GeV is:
\beq
(m_s+m_u)(2)\geq 80 ~{\rm MeV}~~~~\Lrar~~~~ (m_u+m_d)(2)\geq 6.6 ~{\rm MeV}~,
\eeq
if one uses either ChPT and the previous SR analysis for the mass ratios.
Radiative corrections tend to decrease the strengths of these bounds.
Their contributions to the
second moment of the two-point function are (see previous part of the book):
\beq
\Psi^{''}_5(q^2)=\frac{3}{8\pi^2}\frac{(\mb_u+\mb_s)^2}{Q^2}\Bigg{[}1+{11\over
3}\asb+14.179\asb^2+77.368\asb^3\Bigg{]}
\eeq
At this scale, the PT series converges quite well and behaves as:
\beq
{\rm Parton}\Big{[}1+0.45+0.22+0.15\Big{]}~.
\eeq
Including these higher order corrections, the bounds become:
\beq\label{eq: lowest bound}
(m_s+m_u)(2)> (71.4\pm 3.7) ~{\rm MeV}~~~~\Lrar~~~~ (m_u+m_d)(2)>
(5.9\pm 0.3) ~{\rm MeV}~,
\eeq
The bound will be saturated in the extreme limit where the continuum
contribution to the spectral
function is neglected. \\
The quadratic inequality in Eq. (\ref{eq:quad}) results in improved lower
bounds for the quark masses which we show in Fig.~\ref{fig:Fig5b} below.
%%%%%%%%%%%%%%%%%%%
\begin{figure}[htb]
\begin{center}
\includegraphics[width=9cm]{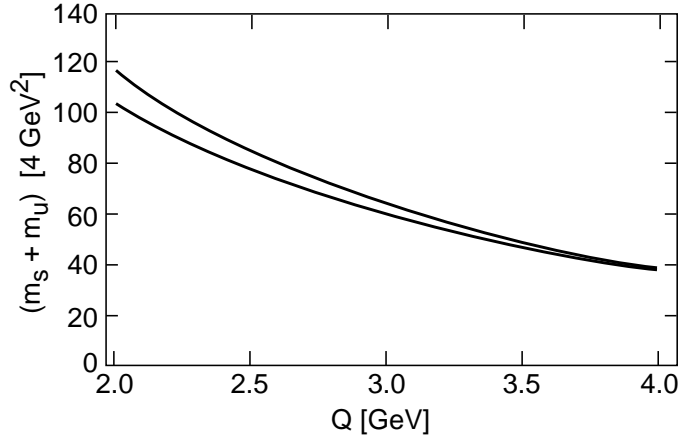}
\caption{The same as in Fig. ~\ref{fig:Fig5a} but from the
quadratic inequality to order $\alpha_s$.}
\label{fig:Fig5b}
\end{center}
\end{figure}
%%%%%%%%%%%%%%%%%%
\nin
The quadratic bound is saturated for a $\delta$--like spectral function
representation of the hadronic continuum of states at an arbitrary
position and with an arbitrary weight. This is certainly less restrictive
than the extreme limit with the full hadronic continuum neglected, and it is
therefore not surprising that the quadratic bound happens to be better than
the ones from
$\Sigma_{N}(Q^2)$ for
$N=0,1,$ and $2$. Notice however that the quadratic bound in
Fig.~\ref{fig:Fig5b} is plotted at higher
$Q$--values than the bound in Fig.~\ref{fig:Fig5a}. This is due to the fact
that the coefficients of the perturbative series in $\als(Q^2)$ become
larger for the higher moments. In Ref~\cite{LEL} it is shown that for the
evaluation of the  quadratic bound $Q\geq 2$ GeV is already a safe
choice.\\
Similar bounds can be obtained for $(m_{u}+m_{d})$ when one considers the
two--point function associated with the divergence of the axial current
\begin{equation}
\partial_{\mu}A^{\mu}(x)=(m_{d}+m_{u}):\!\!{\bar d}(x)i\gamma_{5}u(x)\!\!:.
\end{equation}   The method to derive the bounds is exactly the same as the
one discussed above and therefore we only show, in Fig.~\ref{fig:Fig5c}
below, the results for the corresponding lower bounds  which one obtains
from the quadratic inequality.
%%%%%%%%%%%%%%%%%%%
\begin{figure}[hbt]
\begin{center}
\includegraphics[width=9cm]{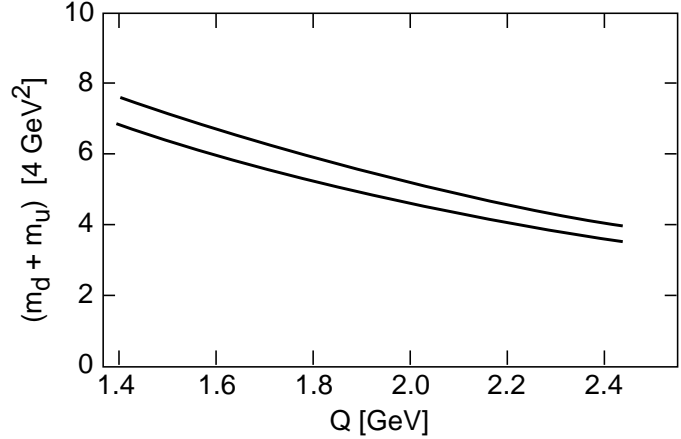}
\caption{Lower bound in  MeV for
$(m_{d}+m_{u})$(2) from the quadratic inequality to order $\alpha_s$.}
\label{fig:Fig5c}
\end{center}
\end{figure}
%%%%%%%%%%%%%%%%%%
At $Q=2$ GeV, one can deduce the lower bounds from the quadratic
inequality:
\beq
(m_s+m_u)(2)> 105 ~{\rm MeV}~,~~~~~~~~~~~~~(m_u+m_d)(2)> 7 ~{\rm MeV}~
\eeq
The convergence of the QCD series is less good here than in the
lowest moment. It behaves as
\cite{KAMBOR}:
\beq
{\rm Parton}\Bigg{[}1+{25\over 3}\asb+61.79\asb^2+517.15\asb^3\Bigg{]}~,
\eeq
which numerically reads:
\beq
{\rm Parton}\Big{[}1+0.83+0.61+0.51\Big{]}~.
\eeq
This leads to the radiatively corrected lower bound to order $\als^3$:
\beq
(m_s+m_u)(2)> (82.7 \pm 13.3)~{\rm MeV}~,~~~~~~~~~~~~~(m_u+m_d)(2)>
(6\pm 1) ~{\rm MeV}~
\eeq
where the error is induced by the truncation of the QCD series which
we have estimated to be about the
contribution of the last known $\als^3$ term of the series
\footnote{In Ref. \cite{STEELE01}, alternative bound has been derived
using a H\"older type inequality. The
lower bound obtained from this method, which is about 4.2 MeV is
weaker than the one obtained previously.}.
  From the previous analysis, and taking into account the uncertainties
induced by the higher order
QCD corrections, {\it the best lower bound} comes from the lowest
inequality and is given in
Eq. (\ref{eq: lowest bound}). The result is summarized in Table
\ref{tab: lowbound}.
\begin{table}[hbt]
\begin{center}
% space before first and after last column: 1.5pc
% space between columns: 3.0pc (twice the above)
\setlength{\tabcolsep}{0.4pc}
% -----------------------------------------------------
% adapted from TeX book, p. 241
%\newlength{\digitwidth} \settowidth{\digitwidth}{\rm 0}
%\catcode`?=\active \def?{\kern\digitwidth}
% -----------------------------------------------------
\caption{Lower bounds on $\overline{m}_{u,d,s}$(2) in MeV}
\label{tab: lowbound}
%\begin{tabular*}{\textwidth}{@{}l@{\extracolsep{\fill}}rrrrr}
\begin{tabular}{c c l }
\hline
%\hline
   && \\
Observables& Sources& Authors\\
&&\\
\hline
&&\\
{ $\overline{m}_u+\overline{m}_d$}&&\\
$6$&$\pi$  &{           LRT97\cite{LEL}, Y97\cite{YNDBOUND} (updated
here to order $\als^3$) }\\
% 7.3&$\pi$&        {   Y97\cite{YNDBOUND} }\\
           6.8&$\la\overline{\psi}\psi\ra$+GMOR& {
DN98\cite{DOSCHSN} (leading order)}\\
%&&\\
%\hline
&&\\
{ $\overline{m}_d-\overline{m}_u$}&&\\
   1.1&$K\pi$&        {  Y97\cite{YNDBOUND} (updated here to order
$\als^3$)        }\\
%&&\\
%\hline
&&\\
{ $\overline{m}_s$}&&\\
           $71.4$&$K$&        { LRT97\cite{LEL} (updated here to order
$\als^3$)        }\\
%         104&$K$&        {  Y97\cite{YND}        }\\
           90&$\la\overline{\psi}\psi\ra$+ChPT& {
DN98\cite{DOSCHSN} (leading order)}\\
&& \\
\hline
\end{tabular}
\end{center}
\end{table}
\nin
%%%%%%%%%%%%%%%%%%%%%%%%%%%%%%%%%%%%%%%%%%%%%%%%%%%%%%%%%
\subsection{Lower bound on the light quark mass-difference from the
scalar sum rule}
As in \cite{SCAL}, one can extract lower bound on the light quark
mass difference
$(m_u-m_d)$ and $(m_u-m_s)$ working with the two-point function
associated to the divergence
of the vector current:
\beq
\partial _\mu V^\mu_{\bar uq}=(m_u-m_q):\bar\psi_u(i)\psi_q:~.
\eeq
The most recent analysis has been done in \cite{YNDBOUND}. We have
updated the result by
including the
$\alpha_s^3$-term. It is given in Table \ref{tab: lowbound}.
\begin{table}[hbt]
\begin{center}
% space before first and after last column: 1.5pc
% space between columns: 3.0pc (twice the above)
\setlength{\tabcolsep}{1.5pc}
% -----------------------------------------------------
% adapted from TeX book, p. 241
%\newlength{\digitwidth} \settowidth{\digitwidth}{\rm 0}
%\catcode`?=\active \def?{\kern\digitwidth}
% -----------------------------------------------------
\caption{Upper bounds on $\overline{m}_{u,d,s}$(2) in  MeV.}
\label{tab: upbound}
%\begin{tabular*}{\textwidth}{@{}l@{\extracolsep{\fill}}rrrrr}
\begin{tabular}{c c l }
\hline
%\hline
   && \\
Observables& Sources& Authors\\
&&\\
\hline
&&\\
{ $\overline{m}_u+\overline{m}_d$}&&\\
           11.4&$\la\overline{\psi}\psi\ra$+GMOR& {
DN98\cite{DOSCHSN} (leading order)}\\
%&&\\
%\hline
&&\\
{ $\overline{m}_s$}&&\\
           148&$\la\overline{\psi}\psi\ra$+ChPT&
DN98\cite{DOSCHSN}  (leading order)\\
            $147\pm 21$ &$e^+e^-+\tau$-decay&        {  SN99\cite{SNMS}
(to order
$\als^3$)      }\\

&& \\
\hline
\end{tabular}
\end{center}
\end{table}
\nin
%%%%%%%%%%%%%%%%%%%%%%%%%%%%%%%%%%%%%%%%%%%%%%%%%%%%%%%%%%%%%%%%%%%%%%%%%
\subsection{Bounds on the sum of light quark masses from the quark
condensate and $e^+e^-\rar
I=0$ hadrons data.} Among the different results in \cite{DOSCHSN}, we
shall use the range of
the chiral
$\la\bar\psi\psi\ra\equiv\la\bar uu\ra\simeq \la\bar dd\ra$ condensate from the
vector form factor of
$D\rar K^* l\nu$.
Using three-point function sum rules, the form factor reads to leading order:
\begin{eqnarray}\label{srv}
V(0) &=& \frac{m_c(m_D+m_{K^*})}{4 f_D f_{K^*} m_D^2 m_{K^*}}
\exp[(m_D^2-m_c^2)\tau_1 + m^2_{K^*} \tau_2]\\
   &\times&\la\bar\psi\psi\ra\Bigg{\{} - 1 + M_0^2(-\frac{\tau_1}{3} +
\frac{m_c^2}{4}
\tau_1^2 +\frac{2 m_c^2- m_c\,m_s}{6}\tau_1\tau_2)\nonumber \\
&&-\frac{16 \pi}{9} \alpha_s \rho  \la\bar\psi\psi\ra (\frac{2 m_c}{9}
\tau_1 \tau_2
-
\frac{m_c^3}{36} \tau_1^3 \nonumber \\
&&- \frac{2 m_c^3-m_c^2 m_s}{36}\tau_1^2 \tau_2
+\frac{-m_c}{9}\tau_1^2 + \frac{2 m_s}{9}\tau_2^2
+\frac{2}{9}m_s\tau_1\tau_2+
\frac{4}{9}\frac{\tau_2}{m_c})\nonumber \\
&& + \frac{e^{m_c^2 \tau_1}}{
\la\bar\psi\psi\ra}\int_0^{s_{20}}ds_2
\int_{s_2+m_c^2}^{s_{10}}ds_1\, \rho_v(s_1,s_2) e^{-s_1 \tau_1 - s_2
\tau_2}\Bigg{\}}
\nonumber \\
{\rm with}&& \rho_v(s_1,s_2) = \frac{3}{4 \pi^2 \, (s_1-s_2)^3} \times
\\ &&\qquad \Big{\{}m_s((s_1+s_2)(s_1-m_c^2) - 2 s_1 s_2) + m_c((s_1 +
s_2) s_2 - 2 s_2
(s_1-m_c^2))\Big{\}}~. \nonumber
\end{eqnarray}
The factor $\rho\simeq 2\sim 3$ expresses the uncertainty in the
factorization of
the four quark condensate. In our numerical analysis,
we start from standard values of the QCD parameters
and  use $f_{K^*} = 0.15~ {\rm GeV} (f_\pi=93.3~\rm{MeV})$.
The value of $f_D\simeq (1.35\pm 0.07)f_\pi$ is consistently
determined  by a two-point function sum rule including radiative
corrections as we shall see
in the next chapter, where the sum rule expression can, e.g., be
found in \cite{SNB}.
The following parameters enter only marginally:
$m_s(1~{\rm GeV}) = (0.15\sim 0.19)~{\rm GeV},~ s_{10} = (5\sim 7)~
{\rm GeV}^2,~ s_{20} = (1.5\sim 2)~{\rm GeV}^2$.~
Using the conservative range of the charm quark mass: $m_c({\rm
pole})$ between 1.29 and 1.55
GeV (the lower limit comes from the estimate in \cite{SNB} and the
upper limit is 1/2 of the
$J/\Psi$ mass), one can deduce the running condensate value at 1 GeV
\cite{DOSCHSN}:
\beq\label{eq: psidosch}
0.6\leq \la\bar\psi\psi\ra/[-225~\rm{MeV}]^3\leq 1.5.
\eeq
This result has been confirmed by the lattice \cite{VALDIKAS}. Using
the GMOR relation:
\beq
2m^2_\pi f^2_\pi=-(m_u+m_d)\la\bar uu+\bar dd\ra +{\cal O}(m^2_q) ~.
\eeq
one can translate the upper bound into a lower bound on the sum of light
quark masses. The lower bound on the chiral condensate can be used in
conjonction with the
positivity of the $m^2_q$ correction in order to give an upper bound
to the quark mass
value. In
this way, one obtains:
\beq
6.8~{\rm MeV}\leq (\bm_u+\bm_d)(2~{\rm GeV})\leq 11.4~{\rm MeV}~.
\eeq
The resulting values are
quoted in Tables
\ref{tab: lowbound} and
\ref{tab: upbound}. We expect that these bounds are satisfied within
the typical 10\% accuracy of the
sum rule approach.\\
   We also show in Table \ref{tab: upbound} the upper bound obtained in
\cite{SNMS} by using
the positivity of the spectral function from the analysis of the
$e^+e^-\rar$ $I=0$ hadrons data
where the determination will be discussed in the next section.
%%%%%%%%%%%%%%%%%%%%%%%%%%%%%%%%%%%%%%%%%%%%%%%%%%%%%%%%%%%%%%%
\section{Sum of light quark masses from pseudoscalar sum rules}
%%%%%%%%%%%%%%%%%%%%%%%%%%%%%%%
\subsection{The (pseudo)scalar Laplace sum rules}
   The Laplace  sum rule for the (pseudo)
scalar two-point correlator reads (see e.g. \cite{43}--\cite{JAMIN2}:
\beq\label{eq: srpseudo}
\int_0^{t_c}dt \exp{(-t\tau)}\frac{1}{\pi}\mbox{Im}\Psi_{(5)}(t)\simeq
(\mb_u\pm \mb_d)^2\frac{3}{8\pi^2}\tau^{-2}
\Bigg{[}\ga 1-\rho_1\dr\ga 1+\delta^{(0)}_{\pm}
\dr+\sum_{n=2}^6\delta^{(n)}_{\pm}\Bigg{]}~,
\eeq
where the indices 5 and +  refer to the pseudoscalar current.
Here, $\tau$ is the Laplace sum rule variable, $t_c$ is the
QCD continuum threshold
and $\mb_i$ is the running mass to
three loops,
\beq
\rho_1\equiv (1+t_c\tau) \exp(-t_c\tau)~.
\eeq
Using the results compiled in previous
Chapter
%\ref{chap: two point function}
, the perturbative QCD corrections read for n flavours:
\bea
\delta^{(0)}_{\pm}&=&\asb\Bigg{[}
\frac{11}{3}-\gamma_1\gamma_E
\Bigg{]}\nnb\\
&+&\asb^2\Bigg{[} \frac{10801}{144}-\frac{39}{2}\zeta(3)
-\ga \frac{65}{24}-\frac{2}{3}\zeta(3)\dr n\nnb\\
&-&\frac{1}{2}\ga 1-\gamma_E^2\dr
\Big{[}\frac{17}{3}\ga 2\gamma_1-\beta_1\dr+2\gamma_2\Big{]}\nnb\\
&+&\ga 3\gamma_E^2-6\gamma_E-\frac{\pi^2}{2}
\dr\frac{\gamma_1}{12}\ga 2\gamma_1-\beta_1\dr\Bigg{]}\nnb\\
\delta^{(2)}_{\pm}&=&-2\tau\Bigg{[} \Big{[}1+
\asb C_F(4+3\gamma_E)\Big{]}\ga
\mb^2_i+\mb^2_j\dr\nnb\\
&\mp& \Big{[}1+\asb C_F\ga 7+
3\gamma_E\dr\Big{]}\mb_i\mb_j\Bigg{]}~,
\eea
where $C_F$=4/3
and $\gamma_E=0.5772...$ is the Euler constant; $\gamma_1,~\gamma_2$
and $\beta_1,~
\beta_2$ are respectively the mass anomalous dimensions
and $\beta$-function coefficients defined in a previous chapter. For
three colours and three flavours, they read:
\beq
\gamma_1=2~,~~~~~~~ \gamma_2=91/12~,~~~~~~~\beta_1=-9/2~,~~~~~~~\beta_2=-8~.
\eeq
In practice, the perturbative correction to the sum rule simplifies as:
\bea
\delta^{(0)}_{\pm}&=&4.82a_s+21.98a_s^2+53.14a_s^3+{\cal
O}(a_s^4)~~~:~~~ a_s\equiv \asb~.
\eea
Introducing
the RGI condensates defined in the previous chapter
, the non perturbative contributions are \cite{BNP}:
\bea
\delta^{(4)}_{\pm}&=&\frac{4\pi^2}{3}
\tau^2\Bigg{[}\frac{1}{4}\overline{\la
\frac{\alpha_s}{\pi}G^2\ra}-\frac{\gamma_1}{\beta_1}\asb\sum_i
\overline{\la m_i\bar\psi_i\psi_i\ra}-
\frac{3}{8\pi^2}\frac{1}{({4\gamma_1+\beta_1})}
\sum_i\mb^4_i\nnb\\
&+&\Big{[}1+\asb C_F\ga\frac{11}{4}+\frac{3}{2}\gamma_E\dr\Big{]}\ga
\overline{\la m_j\bar\psi_j\psi_j\ra}+
\overline{\la m_i\bar\psi_i\psi_i\ra}\dr\nnb\\
&\mp&\Big{[}2+
\asb C_F\ga 7+3\gamma_E\dr\Big{]}\ga
\overline{\la m_i\bar\psi_j\psi_j\ra}+
\overline{\la m_j\bar\psi_i\psi_i\ra}\dr\nnb\\
&-&\frac{3}{2\pi^2}
\Bigg{[}\frac{1}{({4\gamma_1+\beta_1})}\Big{[}
   \frac{\pi}{\alsb}+C_F\ga \frac{11}{4}+
\frac{3}{2}\gamma_E\dr+
\frac{1}{6}\ga 4\gamma_1+\beta_1\dr\nnb\\
&-&\frac{1}{4\gamma_1}
\ga 4\gamma_2+\beta_2\dr~
\Big{]}-\frac{1}{4}\ga 1-2\gamma_E\dr
\Bigg{]}\ga\mb_i^4+\mb_j^4\dr-\frac{3}{2\pi^2}\mb^2_j\mb^2_i\nnb\\
&\pm&\Bigg{[}\frac{1}{({4\gamma_1+\beta_1})}\Big{[}
\frac{2\pi}{\alsb}+\frac{1}{3}\ga 4\gamma_1+\beta_1\dr-\frac{1}{2\gamma_1}
\ga 4\gamma_2+\beta_2\dr
+C_F\ga 7+3\gamma_E\dr
\Big{]}\nnb\\
&+&\gamma_E\Bigg{]}\ga \mb^3_j\mb_i+\mb^3_i\mb_j
\dr\Bigg{]}~,\nnb\\
\delta^{(6)}_{\pm}&=&\mp\frac{8\pi^2}{3}\tau^3\Bigg{[}
\frac{1}{2}\Big{[} m_j\la\bar\psi_i
\sigma^{\mu\nu}\frac{\lambda_a}{2}G^a_{\mu\nu}\psi_i\ra+
m_i\la\bar\psi_j
\sigma^{\mu\nu}\frac{\lambda_a}{2}G^a_{\mu\nu}\psi_j\ra\Big{]}\nnb\\
&-&\frac{16}{27}\pi\alsb\Big{[}\la \bar\psi_j\psi_j\ra^2+
\la \bar\psi_i\psi_i\ra^2\mp 9\la \bar\psi_j\psi_j\ra
\la \bar\psi_i\psi_i\ra\Big{]}\Bigg{]}
\eea
Beyond the SVZ expansion, one can have two contributions:
\begin{itemize}
\item The direct instanton contribution can be obtained from
\cite{SHURYAK} and reads:
\beq
\delta^{inst}_+=\frac{\rho_c^2}{\tau^3}\exp\ga{-r_c}\dr\Big{[}
K_0(r_c)+K_1(r_c)\Big{]}
\eeq
with:
$
r_c\equiv{\rho_c^2/ (2\tau)}$;~ $\rho_c\approx 1/600$ MeV$^{-1}$
being the instanton radius;
$K_i$ is the Mac Donald function. However, one should notice that 
analogous contribution
in the scalar channel leads to some contradictions (\cite{SHURYAK} 
and private communication from Valya
Zakharov).
\item The tachyon gluon mass contribution can be deduced from \cite{CNZ}:
\beq
\delta^{tach}_\pm=-4\asb \lambda^2~,
\eeq
where $(\als/\pi)\lambda^2\simeq -0.06$ GeV$^2$ \cite{CNZ},
\end{itemize} which completes the different QCD contributions to the
two-point correlator.
%\end{document}
%%%%%%%%%%%%%%%%%%%%%%%%%%%%%%%%
\subsection{The $\bar ud$ channel}
  From the experimental side, we
   do not still have a complete measurement of the pseudoscalar
spectral function. In the past
\cite{SNB}, one has introduced the radial excitation $\pi'$ of the
pion using a NWA where the decay
constant has been fixed from chiral symmetry argument \cite{14} and
from the pseudoscalar sum rule
analysis itself \cite{SNP2},\cite{PAVERTR},\cite{SNB}, through the quantity:
\beq
r_\pi\equiv{M^4_{\pi'}f^2_{\pi'}\over m^4_{\pi}f^2_{\pi}}~.
\eeq
Below the QCD continuum $t_c$, the spectral function is
usually saturated by the pion pole and
   its first radial excitation and reads:
\beq
\int_0^{t_c}
dt\exp{(-t\tau)}\frac{1}{\pi}\mbox{Im}\Psi_5(t)\simeq
2m^4_\pi f^2_\pi\exp{(-m_\pi^2 \tau)}
\Bigg{[} 1+r_\pi\exp{\Big{[}\ga m^2_\pi-M^2_{\pi'}
\dr\tau\Big{]}}\Bigg{]}.
\eeq
The theoretical estimate of the spectral function enters through
the not yet measured ratio $r_\pi$. Detailed discussions of the
sum rule analysis can be found in \cite{SNB,SNP,SNP2}. However, this
channel is quite peculiar due to the
Goldstone nature of the pion, where the value of the sum rule scale
($1/\tau$ for Laplace and
$t_c$ for FESR) is relatively large of about 2 GeV$^2$ compared with
the pion mass, where the
duality between QCD and the pion is lost. Hopefully, this paradox can be
cured by the presence of the new $1/q^2$
\cite{ZAKA,CNZ,SNZAK} due to the tachyonic gluon mass, which enlarges
the duality region to lower scale
and then minimizes the role of the higher states into the sum rule.
This na\"\i ve NWA parametrization has
been improved in \cite{LOEWE} by the introduction of threshold effect
and finite width corrections.
Within the advent of ChPT, one has been able to improve the previous
parametrization by imposing
constraints consistent with the chiral symmetry of QCD \cite{BIJP}.
In this way, the spectral function
reads:
\beq
\frac{1}{\pi}\mbox{Im}\Psi_5(t)\simeq
2m^4_\pi f^2_\pi\Big{[}\delta(t-m^2_\pi)+\theta(t-9m^2_\pi)\frac{1}{(16\pi^2
f^2_\pi)^2}\frac{t}{18}\rho^{3\pi}(t)\Big{]}
\eeq
with:
\bea
\rho^{3\pi}(t)&=&\int_{4m^2_\pi}^{(\sqrt{t}-m_\pi)^2}\frac{du}{t}\sqrt{\lambda\ga 
1,{u\over
t},{m^2_\pi\over t}\dr}
\sqrt{1-{4m^2_\pi\over u}}\Bigg{\{}5+\frac{1}{2(t-m^2_\pi)^2}\Big{[}{4\over
3}\big{[}t-3(u-m_\pi^2)\big{]}^2\nnb\\
&&+{8\over 3}\lambda(t,u,m^2_\pi)\ga 1-{4m^2_\pi\over u}\dr+10m^4_\pi\Big{]}+
\frac{1}{(t-m^2_\pi)}\big{[}3((u-m_\pi^2)-t+10m^2_\pi\big{]}\Bigg{\}}~,
\eea
where $\lambda(a,b,c)=a^2+b^2+c^2-2ab-2bc-2ca$ is the usual phase space factor.
Based on this parametrization but including finite width corrections,
a recent re-analysis of this sum rule
   has been given to order $\als^2$ \cite{BIJP}.
Result from the LSR
is, in general, expected to be more reliable than the one from the
FESR due to the presence of
the exponential factor which suppresses the high-energy tail of the
spectral function, though the two
analysis are complementary. In \cite{BIJP}, FESR has been used for
matching by duality the phenomenological
and theoretical parts of the sum rule. This matching has been
achieved in the energy region around 2 GeV$^2$, where the
optimal value of $m_u+m_d$ has been extracted.
In \cite{SNL}, the LSR analysis has been updated by including the
$\alpha_s^3$ correction
obtained in \cite{CHETDPS}. In this way, we get:
\beq
(\mb_u+\mb_d)(2~\mbox{GeV})= (9.3\pm 1.8)~\mbox{MeV}~,
\eeq
where we have converted the original result obtained at the {\it
traditional} 1 GeV to the lattice
choice of scale of 2 GeV through:
\beq
\overline{m}_i(1~{\rm GeV})\simeq  (1.38\pm 0.06)~\overline{m}_i(2~{\rm GeV}),
\eeq
   for running, to order $\alpha_s^3$, the
results from 1 to 2 GeV. This number corresponds to the average value
of the QCD scale
$\Lambda_3\simeq (375\pm 50)$ MeV from PDG \cite{PDG} and
\cite{BETHKE}. Analogous value of $(9.8\pm 1.9)$
MeV for the quark mass
has also been obtained in \cite{PRADES98} to order $\alpha_s^3$ as an
update of the \cite{BIJP}
result. We take as a final result the average from \cite{SNL} and
\cite{PRADES98}:
\beq
(\mb_u+\mb_d)(2~\mbox{GeV})= (9.6\pm 1.8)~\mbox{MeV}~,
\eeq
The inclusion
of the tachyonic gluon mass term reduces this value to \cite{CNZ}:
\beq
\Delta^{tach}(\mb_u+\mb_d)(2~\mbox{GeV})\simeq -0.5~\mbox{MeV}~.
\eeq
As already mentioned, adding to this effect the one of direct
instanton might lead to a double counting
in a sense that they can be alternative ways for parametrizing the
nonperturbative QCD vacuum.
Considering this contribution as
another source of errors, it gives:
\beq
\Delta^{inst}(\mb_u+\mb_d)(2~\mbox{GeV})\simeq -0.5~\mbox{MeV}~.
\eeq
   Therefore, adding different sources of errors, we deduce from the analysis:
\beq
(\mb_u+\mb_d)(2~\mbox{GeV})= (9.6\pm 1.8\pm 0.4 \pm 0.5\pm 0.5)~\mbox{MeV}~,
\eeq
leading to the conservative result for the sum of light quark masses:
\beq\label{eq: sumud}
(\mb_u+\mb_d)(2~\mbox{GeV})= (9.6\pm 2.0)~\mbox{MeV}
\eeq
The first error comes from the SR analysis, the second one comes from
the running mass evolution
and the two last errors come respectively from the (eventual)
tachyonic gluon and direct instanton
contributions. This result is in agreement with previous
determinations \cite{SNB},
\cite{SNP}--\cite{LOEWE}, \cite{PSEU,BRAMON} though
we expect that the errors given there have been underestimated. One
can understand that the
new result is lower than the old result \cite{SNB,SNP} obtained without the
$\alpha_s^2$ and
$\alpha_s^3$ terms as both corrections enter with a positive sign in
the LSR analysis.
However, it is easy to check that the QCD perturbative series
converge quite well in the region
where the optimal result from LSR is obained. Combining the previous
value in Eq. (\ref{eq:
sumud}) with the ChPT mass ratio, one can also deduce:
\beq\label{eq: muchpt}
\mb_s(2~\mbox{GeV})= (117.1\pm 25.4)~\mbox{MeV}~.
\eeq
%%%%%%%%%%%%%%%%%%%%%%%%%%%%%%%%%%%%%%%%%%
\subsection{The $\bar us$ channel and QSSR prediction for the ratio
$m_s/(m_u+m_d)$}
Doing analogous analysis for the kaon channel, one can also derive
the value of the sum $(m_u+m_s)$.
The results obtained from \cite{SNP} updated to order $\alpha_s^3$
and from \cite{DPS99} are shown in Table
\ref{tab: ms} given in \cite{SNL} but updated. We add to the original
errors the one from the tachyonic
gluon (5.5\%), from the direct instanton (5.5\%) and the one due to
the evolution from 1 to 2 GeV (4.4\%),
which altogether increases the original errors by 8.9\%. Therefore,
we deduce the (arithmetic) average from
the kaon channel:
\beq\label{eq: sumus}
\mb_s(2~\mbox{GeV})= (116.0\pm 18.1)~\mbox{MeV}~,
\eeq
One should notice here that, unlike the case of the pion,
the result is less sensitive to the contribution of the higher states
continuum due to the relatively
higher value of $M_K$, though the parametrization of the spectral
function still gives larger errors
than the QCD series.It is
interesting to deduce from Eqs. (\ref{eq: sumud}) and (\ref{eq:
sumus}), the sum rule prediction for the
scale invariant quark mass ratios:
\beq\label{eq: ms/mud}
r_3\equiv\frac{2m_s}{m_u+m_d}\simeq 24.2~,
\eeq
where we expect that the ratio is more precise than the absolute
values due to the cancellation of
the systematics of the SR method. This ratio compares quite well with
the ChPT ratio \cite{14}:
\beq
r_3^{CA}= 24.4\pm 1.5~,
\eeq
and confirms the self-consistency of the pseudoscalar SR approach.
This is a non-trivial test of the SR
method used in this channel and may confirm a posteriori the neglect
of less controlled contributions
like e.g. direct instantons.
\begin{table}[H]
\begin{center}
% space before first and after last column: 1.5pc
% space between columns: 3.0pc (twice the above)
\setlength{\tabcolsep}{0.2pc}
% -----------------------------------------------------
% adapted from TeX book, p. 241
%\newlength{\digitwidth} \settowidth{\digitwidth}{\rm 0}
%\catcode`?=\active \def?{\kern\digitwidth}
% -----------------------------------------------------
\caption{QSSR determinations of $\overline{m}_s$(2) in MeV to order
$\alpha_s^3$.
Some older results have been updated by the inclusion of the higher
order terms. The error
contains the evolution from 1 to 2 GeV. In addition, the errors in
the (pseudo)scalar channels contain
the ones due to the small size instanton and tachyonic gluon mass.
Their quadratic sum increases the original errors by 8.9\%.
The estimated error in the average comes from an arithmetic average
of the different errors.}
\label{tab: ms}
%\begin{tabular*}{\textwidth}{@{}l@{\extracolsep{\fill}}lcl}
\begin{tabular}{l l c l}
%\hline
\hline
                &&& \\
   Channels&$\overline{m}_s$ (2)&Comments& Authors\\
&&&\\
%\hline
\hline
&&&\\
Pion SR + ChPT &$117.1\pm 25.4$&${\cal O}(\als^3)$ &
SN99  \cite{SNL}
Eq. (\ref{eq: muchpt})\\ &&&\\
$\la\bar\psi\psi\ra$+ ChPT &$129.3\pm 23.2$&$N,~B-B^*$ (l.o)&              DN98
\cite{DOSCHSN} Eq. (\ref{eq: msmix})\\
&$117.1\pm 49.0$&$D\rar K^* l\nu$ (l.o)& DN98 \cite{DOSCHSN} Eq.
(\ref{eq: qqmass2})\\
   &&&\\
Kaon SR
                &$119.6\pm 18.4$& updated to ${\cal O}(\als^3)$&
{                 SN89 \cite{SNP,SNB}}\\
                &$112.3\pm 23.2$& ${\cal O}(\als^3)$&               {
DPS99
\cite{DPS99}}\\

						&$116\pm 12.8$&" &
{                 KM01 \cite{KAMBOR}}\\
   %          &$\bf 155\pm 25$& &\bf Largest Range\\
                &&&\\
%\hline
%&&&\\
Scalar SR
                &$148.9\pm 19.2$&${\cal O}(\als^3)$ &               {
CPS97
\cite{CPS97}}\\
                &$103.6\pm 15.4$&" &               {
CFNP97 \cite{COLAPAV}}\\
                &$115.9\pm 24.0$&" &               {
J98 \cite{J98}}\\
                &$115.2\pm 13.0$&" &               {
M99 \cite{M99}}\\

					 &$99\pm 18.3$&" &
{                 JOP01 \cite{OLLER}}\\
%           &$\bf 175\pm 48$& &\bf Largest Range\\
                &&&\\
%\hline
%&&&\\
$                       \tau$-like $\phi$ SR:~  $e^+$-$e^-$  $+\tau$-decay
                &$129.2\pm 25.6$&average: ${\cal O}(\als^3)$ &
{                 SN99
\cite{SNMS}}\\
%                             &$127.5\pm 25.2$&$\Delta_{10}$ &
%{                 }\\
%                                       &$134.8\pm 25.2$&$\Delta_{1\phi}$ &
%{                }\\
%          &$\bf 179\pm 39$& &\bf Largest Range\\
%          &$\leq 200\pm 28$& $R_\phi$+&positivity\\

                &&&\\
%\hline
%&&&\\
$\Delta S=-1$  part of $\tau$-decay
                &$169.5^{+46.7}_{-57}$&${\cal O}(\als^2)$ &
ALEPH99$^{*}$\cite{DAVIERS}\\
                &$144.9\pm 38.4$&"&                            CKP98
\cite{CKP98}\\
                &$114\pm 23$& "&                             PP99 \cite{PP99}\\

					&$125.7\pm 25.4$&"&
KKP00 \cite{KKP00}\\
&$115\pm 21$&" &                              KM01 \cite{KM01}\\

					&$116^{+20}_{- 25}$&" &
CDGHKK01\cite{CDGH01}\\
&&&\\
\hline
&&&\\
\multicolumn{1}{l}
{\bf Average}& $117.4\pm 23.4~$&
\\ &&&\\
\hline
\end{tabular}
\end{center}
{*} Not included in the average.
\end{table}
%%%%%%%%%%%%%%%%%%%%%%%%%%%%%%%%%%%%%%%%%%%%%%%%%%%%%%%%%%%%%%%%%%%
\section{Direct extraction of the chiral condensate $\la\bar uu\ra$}
As mentioned in previous section, the chiral $\bar uu$ condensate can be
extracted directly from the nucleon, $B^*$-$B$ splitting and vector
form factor of $D\rar K^*
l\nu$, which are particularly sensitive to it and to the mixed condensate
$\la\bar\psi\sigma^{\mu\nu}(\lambda_a/2)G^a_{\mu\nu}\psi\ra\equiv
M^2_0\la\bar\psi\psi\ra$
\cite{DOSCHSN}. We have already used the result from the $D\rar K^*
l\nu$ form factor in order to derive upper and lower bounds on
$(m_u+m_d)$. Here, we shall use the
informations from the nucleon and from the $B^*$-$B$ splitting in
order to give a more accurate
estimate.
In the nucleon sum rules \cite{DOSCH}--\cite{IOFFEBAR},
\cite{SNB}, which seem, at first sight, a very good
place for determining $ \la\bar\psi\psi\ra$,  we have two
form factors for which spectral sum rules can be constructed, namely
the form factor $F_1$ which is proportional to the Dirac matrix
$\gamma\,p$ and $F_2$ which is proportional to the unit matrix. In
$F_1$ the four quark condensates play an important role, but these are
not chiral symmetry breaking and are related to the condensate $
\la\bar\psi\psi\ra$ only
by the factorization
hypothesis \cite{SVZ} which is known to be violated by a factor 2-3
\cite{DOSCH,LNT,SNB}.
The form factor $F_2$ is dominated by the condensate $
\la\bar\psi\psi\ra$ and the mixed
condensate $\la\bar{\psi} \sigma G \psi\ra$, such that the baryon mass is
essentially determined by the ratio $M_0^2$ of the two condensates:
\begin{equation}\label{mixed}
M_0^2 = { \la\bar{\psi} \sigma G \psi\ra}/{ \la\bar\psi\psi\ra}
\end{equation}
Therefore from nucleon sum rules one  gets a rather reliable
determination of $M_0^2$ \cite{IOFFEBAR,DOJAMIN}:
\begin{equation} \label{m0}
M_0^2 = (.8\pm .1)~\rm{GeV}^2.
\end{equation}
A sum rule based on the ratio $F_2/F_1$ would in principle be
ideally suited for a determination of $ \la\bar\psi\psi\ra$ but
this sum rule is
completely unstable \cite{DOJAMIN} due to fact that odd parity
baryonic
excitations contribute with different signs to the spectral
functions
of $F_1$ and $F_2$.
In the correlators of heavy mesons ($B,B^*$ and $D,D^*$) the chiral
condensate gives a significant direct contribution in contrast to
the light meson sum rules \cite{SNB}, since, here, it is multiplied by the
heavy quark mass. However, the dominant contribution to the meson mass
comes from the heavy quark mass and therefore a change of a factor
two
in the value of $ \la\bar\psi\psi\ra$ leads only to a negligible
shift of the mass. However, from the $B$-$B^*$ splitting one gets
a precise determination of the mixed condensate
$\la\bar{\psi} \sigma G \psi\ra $
with the value \cite{SNMIXED}
\begin{equation}\label{mixed2}
\la\bar{\psi} \sigma G \psi\ra =  -(9\pm 1)\times 10^{-3}~\rm{GeV^5}~,
\end{equation}
which combined with the value of $M_0^2$ given in Eq. (\ref{m0}) gives
our first result for the value of $ \la\bar\psi\psi\ra$
at the nucleon scale:
\begin{equation}\label{result1}
   \la\bar\psi\psi\ra(M_N)= -[(225 \pm 9\pm 9)~\rm{MeV}]^3~,
\end{equation}
where the last error is our estimate of the systematics and higher 
order contributions.
Using the GMOR relation,
one can translate the previous result into a prediction on the sum of
light quark masses. The
resulting value is:
\cite{DOSCHSN}:
\beq\label{eq: qqmass}
(\mb_u+\mb_d)(2~\mbox{GeV})= (10.6\pm 1.8\pm 0.5)~\mbox{MeV}~,
\eeq
where we have added the second error due to the quark mass evolution.
Combining this value
with the ChPT mass ratio, one obtains:
\beq\label{eq: msmix}
\mb_s(2~\mbox{GeV})\simeq 129.3\pm 23.2~\mbox{MeV}~.
\eeq
Alternatively, one can use the central value of the range given in
Eq. (\ref{eq: psidosch}) in order
to deduce the estimate:
\beq\label{eq: qqmass2}
(\mb_u+\mb_d)(2~\mbox{GeV})= (9.6\pm 4\pm 0.4)~\mbox{MeV}~~~~\Lrar~~~~
\mb_s(2~\mbox{GeV})\simeq (117.1\pm 49.0)~\mbox{MeV}~.
\eeq
The results for $m_s$ are shown in Table \ref{tab: ms}.
%%%%%%%%%%%%%%%%%%%%%%%%%%%%%%%%%%%%%%%%
\section{Final estimate of $(m_u+m_d)$ from QSSR and consequences on
$m_u,~m_d$ and $m_s$}
One can also notice the
impressive agreement of the previous results from pseudoscalar and
from the other channels.
As the two results in Eqs. (\ref{eq: sumud}), (\ref{eq: qqmass}) and
(\ref{eq: qqmass2}) come from
completely independent analysis, we can take their geometric average and
deduce {\it the final value from QSSR}:
\beq\label{eq: finsumd}
(\mb_u+\mb_d)(2~\mbox{GeV})= (10.1\pm 1.3\pm 1.3)~\mbox{MeV}~,
\eeq
where the last error is our estimate of the systematics.
One can combine this result with the one for the light quark mass
ratios from ChPT
\cite{14}:
\label{eq: chpt}
\bea
r_2^{CA}\equiv\frac{m_u}{m_d}= 0.553\pm 0.043~,~~~~~~~~~~~~~~
r_3^{CA}\equiv\frac{2m_s}{(m_d+m_u)}= 24.4\pm 1.5~.
\eea
Therefore, one can deduce the running masses at 2 GeV:
\beq\label{eq: mumdrun}
\mb_u(2)= (3.6\pm 0.6)~\mbox{MeV}~,~~~~~~~~~
\mb_d(2)= (6.5\pm 1.2)~\mbox{MeV}~,~~~~~~~~~
\mb_s(2)= (123.2\pm 23.2)~\mbox{MeV}~.
\eeq
Alternatively, we can use the relation between the invariant mass
$\hat m_q$ and running mass $\mb_q(2)$
to order
$\alpha_s^3$ in order to get:
\beq\label{eq: inv}
\hat m_q=(1.14\pm 0.05)~\mb_q(2)~,
\eeq
for $\Lambda_3=(375\pm 50)$ MeV.
Therefore, one can deduce the invariant masses:
\beq\label{eq: mumdinv}
\hat m_u= (4.1\pm 0.7)~\mbox{MeV}~,~~~~~~~~~
\hat m_d= (7.4\pm 1.4)~\mbox{MeV}~,~~~~~~~~~
\hat m_s= (140.4\pm 26.4)~\mbox{MeV}~.
\eeq
%%%%%%%%%%%%%%%%%%%%%%%%%%%%%%%%%%%%%%%%%%%%%%%%%%%%%%%%%%%%%%%%
\section{Light quark mass from the scalar sum rules}
As can be seen from Eq. (\ref{eq: srpseudo}), one can also (in principle) use
the isovector--scalar sum rule for extracting
the quark mass-differences $(m_d-m_u)$ and $(m_s-m_u)$, and the 
isoscalar--scalar sum rules
for extracting the sum $(m_d+m_u)$.
%%%%%%%%%%%%%%%%%%%%%%%%%%%%%%%%%%
\subsection{The scalar $\bar ud$ channel}
In the isovector channel, the analysis relies heavily on the less controlled
nature of the
$a_0(980)$ \cite{SNB,SNP,SNP2,SCAL} which has been speculated to be a
four-quark state
\cite{218}. However, it appears that its $\bar qq$ nature is favoured by
the present data \cite{MONT}, and further tests are needed for confirming its
real $\bar qq$ assignement.\\
In the $I=0$ channel, the
situation of the
$\pi$-$\pi$  continuum is much more involved due to the
possible gluonium nature of the low mass and wide $\sigma$ meson
\cite{VENEZIA,HAD99,MONT}, which couples strongly to $\pi\pi$ and then
can be missed in the quenched lattice calculation of scalar gluonia states.\\
Assuming that these previous states are quarkonia states, bounds on 
the quark mass-difference
and sum of quark masses have been derived in 
\cite{SCAL,LEL,YNDBOUND}, while an estimate
of the sum of the quark masses has been recently derived in 
\cite{CP01}. However, in view
of the hadronic uncertainties, we expect that the results from the 
pseudoscalar channels
are much more reliable than the ones obtained from the scalar channel. Instead,
we think that it is more useful to use these sum rules the other way
around. Using the values of the quark masses from the pseudoscalar sum rules
and their ratio from ChPT, one can extract their decay constants which are
useful for testing
the $\bar qq$ nature of the scalar resonances \cite{SNB,HAD99} (we shall come
back to this point in the next
section). The agreement of the values of the quark masses
from the isovector scalar channel with the ones from the
pseudoscalar channel can be interpreted as a strong indication for
the $\bar qq$ nature of the
$a_0(980)$. In the isoscalar channel, the value of the sum of light 
quark masses
obtained recently in \cite{CP01}, though slightly lower, agrees 
within the errors with the one from the
pseudoscalar channel. This result supports the maximal 
quarkonium-gluonium scheme for the broad low
mass $\sigma$ and narrow $f_0(980)$ meson: the narrowness of the 
$f_0$ is due to a destructive
interference, while the broad nature of the $\sigma$ is due to a 
contructive interference allowings its
strong coupling with
$\pi\pi$. These features are very important for the scalar meson 
phenomenology, and need to be tested
further.
\subsection{The scalar $\bar us$ channel}
%%%%%%%%%%%%%%%%%%%%%%%%%%%%%%%%%%%%%%%%%%%
Here, the analysis is mostly affected by the parametrization
of the $K\pi$ phase shift data, which strongly affects the resulting
value of the strange quark mass
as can be seen from the different determinations given in the Table
\ref{tab: ms}.
%%%%%%%%%%%%%%%%%%%%%%%%%%%%%%%%%%%%%%%%%%
\section{Light quark mass-difference from \boldmath$(M_{K^+}-M_{K^0})_{QCD}$}
The mass difference $(m_d-m_u)$ can be related to the QCD part of the kaon mass
difference $(M_{K^+}-M_{K^0})_{QCD}$
from the current algebra relation \cite{14}:
\beq
r_2^{CA}\equiv\frac{(m_d-m_u)}{(m_d+m_u)}=
\frac{m_\pi^2}{M_K^2}\frac{(M^2_{K^0}-M^2_{K^+})_{QCD}}{M_K^2-m^2_\pi}
\frac{m^2_s-\hat{m}^2}{(m_u+m_d)^2}=(0.52\pm 0.05)10^{-3}(r_3^2-1),
\eeq
where $2\hat{m}=m_u+m_d$; the QCD part of the $K^+-K^0$ mass-difference
comes from the estimate
of the electromagnetic term using the
Dashen theorem including next-to-leading chiral corrections \cite{BIJP}.
Using the sum rule prediction of $r_3$ from the ratio of $(m_u+m_d)$
in Eq. (\ref{eq: finsumd})
with the average value of $m_s$ in Table \ref{tab: ms} or the ChPT 
ratio given in
the previous section, one
can deduce to order $\als^3$:
\beq\label{eq: massdif}
(\overline{m}_d-\overline{m}_u)\ga{2~\mbox{GeV}}\dr=
(2.8\pm 0.6)~\mbox{MeV}.
\eeq
Analogous result has been obtained from the heavy-light meson
mass-differences
\cite{ELETSKY}. We shall come back to the values of these masses at
the end of this chapter.
%%%%%%%%%%%%%%%%%%%%%%%%%%%%%%%%%%%%%%%%%%%%%%%%%%%%%%%%%%%%%%%%%%%%%%%%%%%%%%%
\section{The strange quark mass from \boldmath$e^+e^-$ and 
\boldmath$\tau$ decays}
%%%%%%%%%%%%%%%%%%%%%%%%%%%%%%%%%
\subsection{$e^+e^-\rar I=0$ hadrons data and the $\phi$-meson channel}
Its extraction from the vector channel has been done in
\cite{REIND,SNB} and more recently in \cite{SNMS}, while its estimate
from an improved
Gell-Mann-Okubo mass formula, including the quadratic mass corrections,
has been done in \cite{GMONARISON,BROAD,SNB}. More recently, the vector
channel has been reanalysed in \cite{SNMS}
using a $\tau$-like inclusive
decay sum rule in a modern version of the Das-Mathur-Okubo (DMO) sum
rule \cite{17}
discussed in previous chapter.
   The analysis in this vector channel is
interesting as we have complete data from $e^+e^-$ in this channel,
which is not
the case of (pseudo) scalar channels where some theoretical inputs
related to the realization of chiral symmetry have to be
used in the parametrization of spectral function.
One can combine the $e^+e^-\rar I=0,~1$ hadrons and the rotated
recent $\Delta S=0$
component of the $\tau$-decay data in
order to extract $m_s$. Unlike previous sum rules, one has the
advantage to have a
complete measurement of the spectral function in the region covered
by the analysis.
We shall work with:
   \bea
R_{\tau,\phi}\equiv\frac{3|V_{ud}|^2}{2\pi\alpha^2}S_{EW}\int_0^{M^2_\tau}
ds\ga 1-\frac{s}{M^2_\tau}\dr^2\nnb\ga
1+\frac{2s}{M^2_\tau}\dr\frac{s}{M^2_\tau}
\sigma_{e^+e^-\rar \phi,\phi',...}~,
\eea
and the $SU(3)$-breaking combinations \cite{SNMS}:
\beq
\Delta_{1\phi}\equiv R_{\tau,1}-R_{\tau,\phi},
~~~~\Delta_{10}\equiv R_{\tau,1}-3R_{\tau,0}~,
\eeq
which vanish in the $SU(3)$ symmetry limit;
$\Delta_{10}$ involves the difference of the isoscalar ($R_{\tau,0}$)
and isovector ($R_{\tau,1}$)
sum rules \`a la DMO. The PT series converges quite well at
the optimization scale of about 1.6 GeV \cite{SNMS}. E.g, normalized
to $\overline{m}^2_s$, one has:
\bea
\Delta_{1\phi}&\simeq& -12\frac{\mb_s^2}{M^2_\tau}\Big{\{}
1+\frac{13}{3}a_s+30.4a_s^2   +(173.4\pm
109.2)a_s^3\Big{\}}\nnb\\
&&+36{\mb_s^4\over M^2_\tau}-36\als^2\frac{\la m_s\bar ss-m_d\bar
dd\ra}{M^4_\tau}~.
\eea
The different combinations $\Delta_{1\phi}$ and $\Delta_{10}$ have
the advantage to be free (to leading
order) from flavour blind combinations like the tachyonic gluon mass
and instanton contributions. We
have checked using the result in \cite{CNZ} that, to non-leading in
$m_s^2$, the tachyonic gluon contribution
is also negligible. It has been argued in \cite{MALTMAN} that
$\Delta_{10}$ can be affected by large $SU(2)$ breakings. This claim
has been tested  using some other
sum rules not affected by these terms \cite{SNMS} but has not been
confirmed. The average from different
combinations is given in Table \ref{tab: ms}. An
upper bound deduced from the positivity of $R_{\tau,\phi}$ is also
given in Table \ref{tab: upbound}.
%%%%%%%%%%%%%%%%%%%%%%%%
\subsection{Tau decays}
Like in the case of $e^+e^-$, one can use tau decays for extracting
the value of $m_s$. However, data
from $\tau$ decays are more accurate than the one from $e^+e^-$. A
suitable combination of sum rule sensitive to leading order to the
$SU(3)$ breaking parameter is needed. It
is easy to construct a such combination which is very similar to the
one for $e^+e^-$. One can work with
the DMO-like sum rule involving the difference between
the $\Delta S=0$ and $\Delta S=-1$ processes \cite{DAVIERS}--\cite{CDGH01}:
\beq
\delta R_\tau^{kl}\equiv {R_{\tau,V+A}^{kl}\over
|V_{ud}|^2}-{R_{\tau,S}^{kl}\over
|V_{us}|^2}=3S_{EW}\sum_{D\geq
2}\Big{\{}\delta_{ud}^{kl(D)}-\delta_{us}^{kl(D)}\Big{\}}~,
\eeq
where the moments are defined as:
\beq
R_{\tau}^{kl}\equiv\int_0^{M^2_\tau}
ds\ga 1-\frac{s}{M^2_\tau}\dr^k\ga \frac{s}{M^2_\tau}\dr^l{dR_\tau\over ds}~,
\eeq
with $R_{\tau}^{00}\equiv R_{\tau}$ is the usual $\tau$-hadronic width. The QCD
expression reads:
\beq
\delta R_\tau^{kl}\simeq 24 S_{EW}\Bigg{\{}{\mb_s^2\over
M^2_\tau}\Delta^{(2)}_{kl}
-2\pi^2{\la m_s\bar ss-m_d\bar dd\ra\over M^4_\tau}\Delta^{(4)}_{kl}\Bigg{\}}~,
\eeq
where $\Delta^{(D)}_{kl}$ are perturbative coefficients known to
order $\als^2$:
\beq
\Delta^{(D)}_{kl}\equiv
\frac{1}{4}\Big{\{}3\Delta^{(D)}_{kl}\big{|}_{L+T}+\Delta^{(D)}_{kl}\big{|}_L\Big{\}}~,
\eeq
where the indices $T$ and $L$ refer to the tranverse and longitudinal
parts of the
spectral functions. For $D=2$, the $L$ piece converges quite badly while the
$L+T$ converge quite well such that the combination has can still an
acceptable convergence. For the lowest moments, one has:
\bea
\Delta^{(2)}_{00}&=&0.973+0.481+0.372+0.337+...\nnb\\
\Delta^{(2)}_{10}&=&1.039+0.558+0.482+0.477+...\nnb\\
\Delta^{(2)}_{20}&=&1.115+0.643+0.608+0.647+...
\eea
The authors advocate that though the convergence is quite bad, the 
behaviour of the series is
typical for an asymptotic series close to their point of minimum 
snesistivity. Therefore, the
mathematical procedure for doing  a reasonable estimate of the series 
is to truncate the expansion
where the terms reach their minimum value. However, the estimate of 
the errors is still arbitrary.
The authors assume that the error is given by the last term of the 
series. The result of the
analysis is given in Table \ref{tab: ms}. The  different numbers given
in the table reflects the difference of methods used to get $m_s$ but
the results are consistent
each others within the errors. Like in the case of the $e^+e^-$
DMO-like sum rule, the combination
used here is not affected to leading order by flavour blind
contribution like the tachyonic gluon
and instanton contribution. We have checked \cite{CNZ} that the
contribution of the tachyonic
gluon to order $m^2_s\alpha_s\lambda^2/M^2_\tau$ gives a tiny
correction and does not affect the
estimate done without the inclusion of this term.

%%%%%%%%%%%%%%%%%%%%%%%%%%%%%%%%%%%%%%%%%%%%%%%%%%%%%%%%%%%
\subsection{Summary for the estimate of light quark masses}\label{sec: lsum}
Here, we summarize the results from the previous analysis:
\begin{itemize}
\item The sum $(\bm_u+\bm_d)$ of the running up and down quark masses
from the pion sum rules is given in
Eq. (\ref{eq: sumud}), while the one of the strange quark mass from
the kaon channel is given in
Eq. (\ref{eq: sumus}). Their values lead to the pseudoscalar sum
rules prediction for the mass ratio in Eq.
(\ref{eq: ms/mud}) which agrees nicely with the ChPT mass ratio.
\item The sum $(\bm_u+\bm_d)$ of the running up and down quark masses
averaged from the pseudoscalar sum rule and
from a direct extraction of the chiral condensate $\la\bar uu\ra$
obtained from a global fit of the nucleon,
$B^*-B$ mass-splitting and the vector part of the $D^*\rar K^*l\nu$
form factor is given in Eq.
(\ref{eq: mumdrun}) and reads for $\Lambda_3=(375\pm 50)$ MeV:
\beq
(\bm_u+\bm_d)(2~{\rm GeV})=(10.1\pm 1.8)~{\rm MeV}~,
\eeq
implying with the help of the ChPT mass ratio $m_u/m_d$, the value:
\beq
\bm_u(2~{\rm GeV})=(3.6\pm 0.6)~{\rm MeV}~,~~~~~~~~~~\bm_d(2~{\rm
GeV})=(6.5\pm 1.2)~{\rm MeV}~,
\eeq
which leads to the invariant mass in Eq. (\ref{eq: mumdinv}):
\beq
\hat m_u=(4.1\pm 0.7)~{\rm MeV}~,~~~~~~~~~~\hat m_d=(7.4\pm 1.4)~{\rm MeV}~,
\eeq
\item We have combined the result in Eq. (\ref{eq: finsumd}) with the
sum rule prediction for $m_s/(m_u+m_d)$
in order to deduce the quark mass-difference $(m_d-m_u)$ from the QCD
part of the $K^0-K^+$ mass-difference.
We obtain the result in Eq. (\ref{eq: massdif}):
\beq
(\bm_d-\bm_u)(2~{\rm GeV})=(2.8\pm 0.6)~{\rm MeV}~.
\eeq
This result indeed agrees with the one taking the difference of the
mass given previously. The fact that
$(m_u+m_d)\not= (m_d-m_u)$ disfavours the possibility to have $m_u=0$.
\item  We give in Table \ref{tab: ms} the different sum rules
determinations of $m_s$.  The results from the
pion SR and $\la\bar\psi\psi\ra$ come from the determination of
$(m_u+m_d)$ to which we have added the ChPT
contraint on $m_s/(m_u+m_d)$. One can see from this table that
different determinations are in good agreement
each others. Doing an average of these different results, we obtain:
\beq
\bm_s(2~{\rm GeV})= (117.4\pm 23.4)~{\rm MeV}~~~~\Lrar~~~~ \hat
m_s=(133.8\pm 27.3)~{\rm
MeV}~.
\eeq
Aware on the possible correlations between these estimates, we have
estimated the error as an arithmetic average
which is about 10\% as generally expected for the systematics of the
SR approach.
\end{itemize}
It is informative to compare the above results with the average of
different quenched and unquenched lattice
values
\cite{LUBICZ}:
\bea
&&\bm_{ud}(2~{\rm GeV})\approx \frac{1}{2}(\bm_u+\bm_d)(2~{\rm
GeV})=(4.5\pm 0.6\pm 0.8)~{\rm MeV}~,\nnb\\
&&\bm_s(2~{\rm GeV})= (110\pm 15\pm 20)~{\rm MeV}~,
\eea
where the last error is an estimate of the quenching error. We show
in the Table \ref{tab: mqunq}
   a compilation of the lattice unquenched results including comments
on the lattice characterisitcs
(action, lattice spacing $a$, $\beta$). Also shown is the ratio over
$m_s/m_{ud}$ and quenched (quen) over
unquenched (unq) results.
%____________________________________________________________________
\begin{table}[hbt]
\caption{Simulation details and physical results of unquenched lattice
calculations of light quark masses from
\cite{LUBICZ}, where original references are quoted.}
\label{tab: mqunq}
\renewcommand{\tabcolsep}{0.3pc} % enlarge column spacing
\renewcommand{\arraystretch}{1.2} % enlarge line spacing
\begin{tabular}{lcccclccc}
\hline
\\
& Action & $a^{-1}$[GeV] & $\#_{(\beta,K_{sea})}$ & $Z_m$ &
\multicolumn{2}{c}{$\bm_s(2)$} & ${m_s\over m_{ud}}$ &
$\frac{\bm_s^{\mbox{quen}}}{\bm_s^{\mbox{unq}}}$
\\
\\ \hline
\\
SESAM 98  & Wilson & $ 2.3  $ & 4 &  PT
        & $\, $ 151(30) &$(m_{K,\phi})$& 55(12) & 1.10(24)
        \\[8pt]
MILC 99 & Fatlink & $ 1.9  $ & 1 &  PT
        & $\begin{array}{l} 113(11) \\ 125(9) \end{array}$
        & $\begin{array}{c} (m_K) \\ (m_\phi) \end{array}$ & 22(4) & 1.08(13)
        \\[16pt]
APE 00 & Wilson & $ 2.6  $ & 2 &  NP-RI
        & $\begin{array}{l} 112(15) \\ 108(26) \end{array}$
        & $\begin{array}{c} (m_K) \\ (m_\phi) \end{array}$ & 26(2) & 1.09(20)
        \\[16pt]
CP-PACS 00  & MF-Clover & $a \to 0$ & 12 & PT
        & $\begin{array}{l} 88^{+4}_{-6} \\ 90^{+5}_{-11} \end{array}$
        & $\begin{array}{c} (m_K) \\ (m_\phi)\end{array}$ & 26(2) &1.25(7)
        \\[16pt]
JLQCD 00 & NP-Clover & $ 2.0  $ & 5 & PT
        & $\begin{array}{l} 94(2)^\dag \\ 88(3)^\ddag \\ 109(4)^\dag \\
        102(6)^\ddag \end{array}$
        & $\begin{array}{c} (m_K) \\ \\ (m_\phi)\end{array}$ & --- & ---
        \\[16pt] \hspace{-0.3truecm}
$\begin{array}{l} \mbox{\rm QCDSF +} \\ \mbox{\rm UKQCD 00 }
        \end{array}$ & NP-Clover & $ 2.0  $ & 6 &  PT & $\, $ 90(5)
        & $(m_K)$ & 26(2) & --- \\\\
\hline
\end{tabular}\\[2pt]
$^\dag$ From vector WI; $^\ddag$ from axial WI. The errors on the ratios
$m_s/m_{ud}$ and
$\bm_s^{\mbox{quen}}/\bm_s^{\mbox{unq}}$ are
estimates based on the original data.
\vspace*{-.3cm}
\end{table}
%___________________________________________________________________________

%%%%%%%%%%%%%%%%%%%%%%%%%%%%%%%%%%%%%%%%%%%%%%%%%%
\section{Decay constants of light (pseudo)scalar mesons}
%%%%%%%%%%%%%%%%%%%%%%%%%%%%%%%%%
\subsection{Pseudoscalar mesons}
\begin{figure}[hbt]
\begin{center}
\includegraphics[width=9cm]{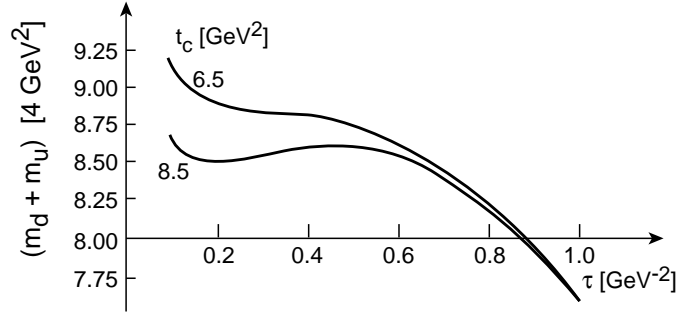}
\caption{LSR analysis of the ratio $r_\pi\equiv 
M^4_{\pi'}f_{\pi'}^2/M^4_{\pi}f_{\pi}^2$. For a
given value $r_\pi=9.5$, we show value of $(\bm_d+\bm_u)(2)$ for two 
values of the QCD continuum
$t_c$.}
\label{fig: rpi}
\end{center}
\end{figure}

Due to the Goldstone nature of the pion and kaon, we have seen that
their radial excitations play an essential
r\^ole in the sum rule. This unusual property allows a determination
of the radial excitation parameters.
In the strange quark channels, an update of the results in
\cite{SNMS,SNP2,SNP,SNB,PAVERTR} gives:
\beq
r_K\equiv M^4_{K'}f_{K'}^2/M^4_{K}f_{K}^2 \simeq 9.5\pm 2.5\simeq r_\pi~,
\eeq
where $r_\pi$ has been defined previously. The optimal value has been
obtained at the LSR
scale $\tau\approx$ GeV$^{-2}$ and $t_c\simeq 4.5-6.5$ GeV$^2$ as 
shown in Fig. \ref{fig: rpi}.
%%%%%%%%%%%%%%%%%%%
%%%%%%%%%%%%%%%%%%
  This
result implies for
$\pi'(1.3)$ and $K'(1.46)$:
\beq\label{eq: fpseu}
f_{\pi'}\simeq (3.3\pm 0.6)~{\rm
MeV}~,~~~~~~~~~~~~~~~~~~~~f_{K'}\simeq (39.8\pm 7.0)~{\rm
MeV}~.
\eeq
   It is
easy to notice that the result satisfies the relation:
\beq
{f_{K'}\over f_{\pi'}}\approx {M^2_K\over m^2_\pi}\approx {m_s\over m_d}~,
\eeq
as expected from chiral symmetry arguments.
%%%%%%%%%%%%%%%%%%%%%%%%%%%%%%%%%%
\subsection{Scalar mesons}
\begin{figure}[hbt]
\begin{center}
\includegraphics[width=9cm]{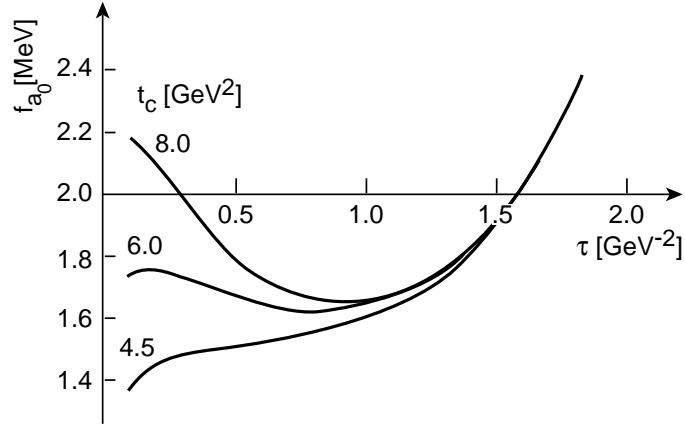}
\caption{LSR analysis of the decay constant $f_{a_0}$ of the 
$a_0(.98)$ meson normalized as $f_\pi=92.4$ MeV.
We use $(\bm_d-\bm_u)(2)=2.8$ MeV.}
\label{fig: fa0}
\end{center}
\end{figure}

\begin{figure}[hbt]
\begin{center}
\includegraphics[width=9cm]{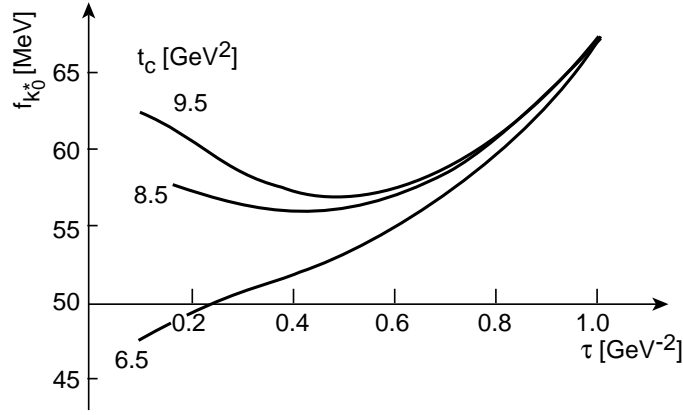}
\caption{LSR analysis of the decay constant $f_{K*_0}$ of the 
$K^*_0(1.43)$ meson normalized as $f_\pi=92.4$
MeV. We use $\bm_s(2)=117.4$ MeV.}
\label{fig: fK*0}
\end{center}
\end{figure}

We expect that the scalar channel is more useful for giving the decay
constants of the mesons
which are not well known rather than predicting the value of the
quark masses. Such a programme has been
initiated in \cite{SNP,SNP2,SNB}. Since then, the estimate of the
decay constants has not mainly changed. The analysis is shown in 
Figs. \ref{fig: fa0} and \ref{fig: fK*0}.
%%%%%%%%%%%%%%%%%%%
%%%%%%%%%%%%%%%%%%
Recent estimate gives \cite{SNMS}:
\beq\label{eq: fscal}
f_{a_0}=(1.6\pm 0.15\pm 0.35\pm 0.25)~{\rm
MeV}~,~~~~~~~~~~~~f_{K^*_0}\simeq (46.3\pm 2.5\pm 5\pm 5)~{\rm MeV}~,
\eeq
where the errors are due respectively to the choice of $t_c$ from 4.5
to 8 GeV$^2$, the value of the quark
mass-difference obtained previously and the one of $\Lambda_3$. The
decay constants are normalized as:
\beq
\la 0|\pr_\mu V^\mu(x)|a_0\ra=\sqrt{2}f_aM^2_a,
\eeq
corresponding to $f_\pi=92.4$ MeV. We have used the experimental
masses 0.98 GeV and 1.43 GeV in our analysis
\footnote{The masses of the $a_0$ and $K^*_0$ are also nicely
reproduced by the ratio of
moments \cite{SRRY,SNB}.}.  It is also interesting to notice that the
ratio of the decay constants are:
\beq
{f_{K^*_0}\over f_{a_0}}\simeq 29\approx {m_s-m_u\over m_d-m_u}\simeq 40~,
\eeq
as na\"\i vely expected. We are aware that the values of these decay
constants might have been overestimated due to
the eventual proliferations of nearby radial excitations. Therefore, it will
be interesting to have a direct measurement of these decay constants
for testing these predictions. The values
of these decay constants will be given like other meson decay
constants in the table of the next chapter.
%%%%%%%%%%%%%%%%%%%
%%%%%%%%%%%%%%%%%%%%%%%%%%%%%%%%%%%%%%%%%%%%%%%%%%%%
\section{Flavour breakings of the quark condensates}
%%%%%%%%%%%%%%%%%%%%%%%%%%%%%%%%%%%%%%%%%%%%%
\subsection{$SU(3)$ corrections to kaon PCAC}
\begin{figure}[hbt]
\begin{center}
\includegraphics[width=9cm]{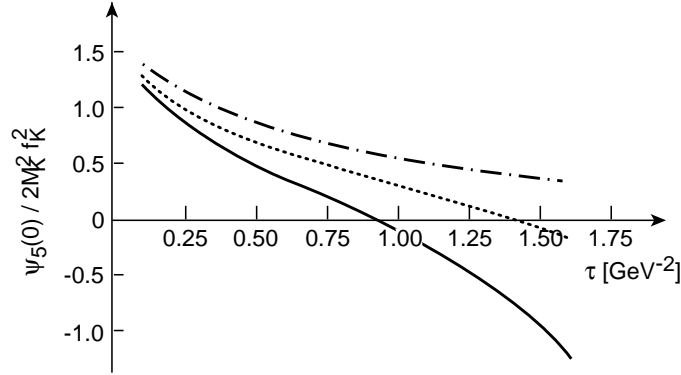}
\caption{LSR analysis of the subtraction constant $\Psi_5(0)$. We use 
$\bm_s(2)=117.4$ MeV, $r_K=9.5$
and $t_c=6$ GeV$^2$.
The curves correspond to different truncations of the PT series: to 
${\cal O}(\alpha_s)$: dotted-dashed;
to ${\cal O}(\alpha_s^2)$: dashed; to ${\cal O}(\alpha_s^3)$: continuous.}
\label{fig: kpcac}
\end{center}
\end{figure}

Let's remind that the (pseudo)scalar two-point function obeys the 
twice subtracted dispersion relation:
\beq
\Psi_{(5)}(q^2)=\Psi_{(5)}(0)+q^2\Psi'_{(5)}(0)+q^4\int_0^\infty{dt\over 
t^2(t-q^2-i\epsilon)}{\rm Im}
\Psi_{(5)}(t)~.
\eeq
The deviation from kaon PCAC has been firstly studied
in \cite{SNSU3} using the once subtracted
peusdoscalar sum rule of the quantity:
\beq
{\Psi_{(5)}(q^2)-\Psi_{(5)}(0)\over q^2}
\eeq
sensitive to the value of the
value of the correlator at $q^2=0$ \footnote{This sum rule has also 
been used in \cite{SPIN,SNSLOPE}for
estimating the $U(1)_A$ topological suceptibility and its slope, and 
which has been checked on the
lattice \cite{DIGIACOMO}.}. The Ward identity obeyed by the 
(pseudo)scalar two-point function leads to the
low-energy theorem:
\beq\label{eq: condnorm}
\Psi_{(5)}(0)=-(m_i\pm m_j)\la \bar \psi_i\psi_i\pm \bar \psi_j\psi_j\ra~,
\eeq
in terms of the {\it normal ordered condensates.} However, as 
emphasized in different papers
\cite{43,BROAD,JAMIN2,CHETDPS}, $\Psi_{(5)}(0)$ contains a 
perturbative piece which cancels the
mass singularities appearing in the OPE evaluation of $\Psi_{(5)}(q^2)$. This leads 
to the fact that
the quark condensate entering in Eq. \ref{eq: condnorm} are defined 
as a {\it non-normal ordered condensate},
which has a slight dependence on the scale and renormalization 
scheme. This mass correction effect is only
quantitatively relevant for the $\bar us$ channel but not for the 
$\bar ud$ one. To order
$\alpha_s^3$ for the perturbative term and to leading order for the 
condensates, the
(pseudo)scalar sum rule for the $\bar us$ channel reads, by 
neglecting the up quark mass:
\bea
&&\int_0^{t_c}\frac{dt}{t} \exp{(-t\tau)}
\frac{1}{\pi}\mbox{Im}\Psi_{(5)}(t)\simeq\Psi_{(5)}(0)\nnb\\
&&+(\mb_u\pm\mb_s)^2\frac{3}{8\pi^2}\tau^{-1}
\Bigg{\{}\ga 1-\rho_0\dr\Big{[}1+
6.82\asb+
58.55\asb^2+537.6\asb^3\Big{]}\nnb\\
&&+3.15\mb^2_s\tau\Big{[}1+3.32\asb\Big{]}\nnb\\
&&-\Bigg{[} \frac{\pi}{3}\la \als G^2\ra-\frac{8\pi^2}{3}
\Big{[}\ga \mb_s-{\mb_u\over 2}\dr\la\bar uu\ra\pm (u\leftrightarrow s)
\Big{]}\Bigg{]}\tau^2\nnb\\
&&+\frac{1}{2}\ga 2\mp 9\dr\ga\frac{128}{81}\dr\pi^3\rho\als\la\bar
uu\ra^2\tau^3
\Bigg{\}},
\eea
where we have neglected the $SU(3)$ breaking for the four-quark 
condensates. This assumption does not
however affects the analysis due to the small contribution of this 
operator at the optimization
scale.
The analysis is shown in Fig. \ref{fig: kpcac}.
%%%%%%%%%%%%%%%%%%%
%%%%%%%%%%%%%%%%%%
Examining the different curves, on can notice that they deviate 
notably from the
kaon PCAC prediction:
\beq
\Psi_5(0)\simeq 2M^2_Kf^2_K~,
\eeq
threfore confirming the early findings in \cite{SNSU3}.
The
LSR indicates a slight stability point at $\tau\approx (0.50\sim 
0.75)$ GeV$^{-2}$, where:
\beq
\Psi_5(0)\simeq (0.5\pm 0.2)2M^2_Kf^2_K~.
\eeq
  However at this scale,
PT series has a bad convergence:
\bea
{\rm Pert}={\rm Parton}\times \Big{\{}1+2.17\als+5.93\als^2+17.34\als^3\Big{\}}
\simeq{\rm Parton}\times \Big{\{}1 +0.86+0.92+1.06\Big{\}}~,
\eea
which might not be worrysome if one considers that asymptotic series 
close to its
point of {\it minimum sensitivity} can be truncated when its reaches 
the extremum value and
add the last term as truncation error \footnote{A similar argument 
has been used for the extraction
of the strange quark mass from $\tau$-decay data discussed in the 
previous section, where the QCD
series also has a quite bad behaviour.}.
  This convergence might {\it a priori} be improved if one works
  with the combination of sum rules which is less sensitive to
the high-energy behaviour of the spectral function (and then to the 
perturbative contribution)
than the former \cite{PAVERTR,BRAMON,SNP,SNB,SNMS}. The modified sum 
rule reads \cite{SNB} \footnote{Notice that we
have not yet introduced  the QCD continuum into the LHS of the sum rule.}:
\bea
&&\int_0^{\infty}\frac{dt}{t} \exp{(-t\tau)}
\ga 1-t\tau\dr\frac{1}{\pi}\mbox{Im}\Psi_{(5)}(t)\simeq\Psi_{(5)}(0)+
(\mb_u\pm\mb_s)^2\frac{3}{8\pi^2}\tau^{-1}\times\nnb\\
&&\Bigg{\{}2\asb\Big{[}1+18.3\asb+242.2\asb^2\Big{]}
+5.15\mb^2_s\tau\Big{[}1+5.0\asb\Big{]}\nnb\\
&&+2\Big{[} \frac{\pi}{3}\la \als G^2\ra-\frac{8\pi^2}{3}
\mb_s\Big{[}\la\bar uu\ra\mp\frac{1}{2}\la\bar ss\ra
\Big{]}\Big{]}\tau^2+\frac{3}{2}(2\mp 
9)\ga\frac{128}{81}\dr\pi^3\rho\als\la\bar uu\ra ^2\tau^3
\Bigg{\}}~.
\eea
The analysis also leads to a similar result. The LSR has been
also studied recently in \cite{KAMALKAN}, by including threshold effects and
higher mass resonances, which enlarge the region of stability in the 
LSR variable. Within the previous
hadronic parametrization, one obtains:
\beq
\Psi_5(0)\simeq (0.56\pm 0.04\pm 0.15)2M^2_Kf^2_K~,
\eeq
where we have added the error due to our estimate of the truncation 
of the QCD PT series as deduced
from Fig. \ref{fig: kpcac}.
%\end{document}
An alternative estimate is the uses of FESR \cite{PSEU}. 
Parametrizing the subtraction constant
as:
\beq
\Psi_5(0)^u_s=2M_K^2f_K^2(1-\delta_K)~,
\eeq
one has the sum rule \cite{PSEU}:
\beq
\delta_K\simeq \frac{3}{16\pi^2}\frac{\overline{m}_s^2t_c}{f^2_KM^2_K}
\aga 1+\frac{23}{3}a_s + {\cal{O}}(a^2_s)\adr
-r_K\ga\frac{M_K}{M_{K'}}\dr^2~,
\eeq
which gives, after using the {\it correlated values} of the input
parameters \cite{SNP,SNB,SNMS}:
\beq
\delta_K= 0.34^{+0.23}_{-0.17}~,
\eeq
leading to:
\beq
\Psi_5(0)\simeq (0.66\pm 0.20)2M^2_Kf^2_K~,
\eeq
confirming the large violation of kaon PCAC obtained from LSR.
%%%%%%%%%%%%%%%%%%%%%%%%%%%%%%%%%%%%%%%%%%%%%%%%%
\subsection{Subtraction constant from the scalar sum rule}
One can do a similar analysis for the scalar channel.
%%%%%%%%%%%%%%%%%%%
\begin{figure}[hbt]
\begin{center}
\includegraphics[width=8cm]{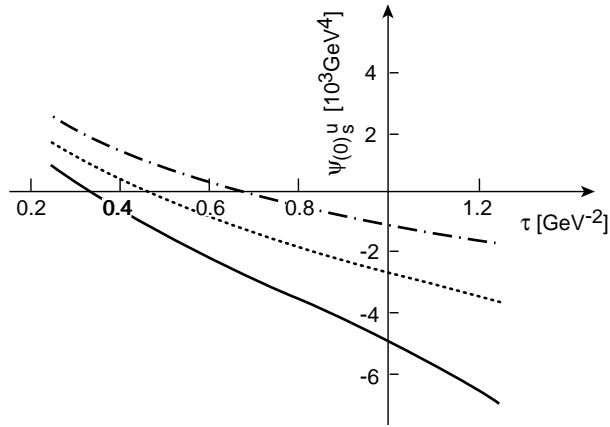}
\caption{LSR analysis of the subtraction constant $\Psi(0)$. We use 
$\bm_s(2)=117.4$
MeV, $f_{K*_0}=46$ MeV and $t_c=6.5$ GeV$^2$. The curves correspond 
to different truncations of
the PT series: to
${\cal O}(\alpha_s)$: dotted-dashed; to ${\cal O}(\alpha_s^2)$: 
dashed; to ${\cal
O}(\alpha_s^3)$: continuous.}
\label{fig: psi0}
\end{center}
\end{figure}
%%%%%%%%%%%%%%%%%%
The analysis from LSR is shown
in Fig. \ref{fig: psi0}. One can also see that there is a slight stability for
$\tau\approx (0.50\sim 0.75)$ GeV$^{-2}$, which gives:
\beq
\Psi(0)\approx -10^{-3}~{\rm GeV}^4~,
\eeq
in agreement with previous results \cite{SNB,SNP,SNP2,PAVERTR}. In 
\cite{KAMALKAN}, using
LSR, a similar result but from a larger range of LSR stability, has 
been obtained  within an
Omn\'es representation for relating the scalar form factor to the 
$K\pi$ phase shift data:
\beq
\Psi(0)\simeq -(1.06\pm 0.21\pm 0.20)10^{-3}~{\rm GeV}^4~,
\eeq
where the last term is our estimate of the error due to the 
truncation of the QCD series.
%We
%show the analysis in Fig. \ref{fig: kamal}.
%%%%%%%%%%%%%%%%%%%
%\begin{figure}[hbt]
%\begin{center}
%\includegraphics[width=8cm]{Figkamal.ps}
%\caption{LSR analysis of the subtraction constant $\Psi(0)$ versus 
the sum rule scale
%using $K\pi$ phase shift data, from
%\cite{KAMALKAN}.}
%\label{fig: kamal}
%\end{center}
%\end{figure}
%%%%%%%%%%%%%%%%%%
%\nin
One can use an alternative approach by working with FESR:
\beq
\Psi(0)^u_s=2M_{K^*_0}^2f_{K^*_0}^2-
\frac{3}{16\pi^2}{\overline{m}_s^2t_c}
\aga 1+\frac{23}{3}a_s + {\cal{O}}(a^2_s)\adr,
\eeq
which gives \cite{SNMS}:
\beq
\Psi(0)^u_s=-\ga 7.8^{+5.5}_{-2.7}\dr 10^{-4}~\mbox{GeV}^4.
\eeq
%%%%%%%%%%%%%%%%%%%%%%%%%%%%%%%%%%%%%%%%%%%%%%%%%
\subsection{${\la \bar ss \ra}/{\la \bar uu \ra}$ from the 
(pseudo)scalar sum rules}
We take the arithmetic average of the previous determinations for our 
final estimate:
\beq
\Psi_5(0)\simeq (0.57\pm 0.19)2M^2_Kf^2_K~,~~~~~~~~~~~~~~~~
\Psi(0)\simeq -(0.92\pm 0.35)10^{-3}~{\rm GeV}^4~,
\eeq
Taking the ratio of the
scalar over the pseudoscalar subtraction constants expressed in terms of
the {\it normal-ordered} condensates, one can deduce:
\beq\label{eq: ssdd5}
{\la \bar ss \ra}/{\la \bar uu \ra} = 0.57\pm 0.12.
\eeq
%%%%%%%%%%%%%%%%%%%%%%%%%%%%%%%%%%%%%%%%%%%%%%%%%%%%%%%%%%%%%%%%%%%%%%
\subsection{${\la \bar ss \ra}/{\la \bar uu \ra}$ from the $B_s$ meson}
One can also extract the flavour breakings of the condensates from a 
sum rule analysis of the $B_s$
and $B^*_s$ masses, which are senstive to the chiral condensate as it 
enters like $m_b\la \bar ss
\ra$ in the OPE of the heavy light meson (see next section). The 
masses of the mesons are found to
decrease linearly with the value of the chiral condensate. Using the 
observed value of the
$B_s$ meson mass $M_{B_s}=5.375$ GeV, one can deduce from Fig 3 of 
\cite{SNMIXED}:
\beq\label{eq: ssbs}
{\la \bar ss \ra}/{\la \bar uu \ra}\simeq 0.75\pm 0.08~,
\eeq
where the error is the expected typical sum rule estimate. The effect 
of the strange quark mass is
less important than this one here, such that the result given in 
\cite{SNMIXED} remains valid though
obtained with slightly different values of $m_s$ and $\Lambda_5$. 
This estimate is expected to be
more reliable than the one from the (pseudo)scalar light mesons, 
which are affected by the bad
convergence of the PT QCD series. Using this value of ratio of the 
condensates in the curve of the
$B^*_s$ mass, one can also predict:
\beq
B^*_s=5.64 ~{\rm GeV}~,
\eeq
which can be tested in $B$ factories.
%%%%%%%%%%%%%%%%%%%%%%%%%%%%%%%%%%%%%%%%%%%%%%%%%%%%%%%%%%%%%%%%%%%%%%%%%%%%
\subsection{Final sum rule estimate of ${\la \bar ss \ra}/{\la \bar uu \ra}$}
Using the previous results, one can deduce that the sum rules from 
the light (pseudo)scalar and
from the $B_s$ meson predict for the {\it normal ordered} condensate ratio:
\beq\label{eq: ssdd}
{\la \bar ss \ra}/{\la \bar uu \ra}\simeq 0.66\pm 0.10~,
\eeq
confirming earlier findings \cite{PAVERTR,SNB,SNP,SNMS} on the large 
flavour breaking of the chiral
condensate.
This number comes from the arithmetic average of the two values in 
Eqs. (\ref{eq: ssdd5}) and
(\ref{eq: ssbs}).
  If
one instead works with the {\it non-normal ordered} condensate, one 
should add to the expression in Eq.
(\ref{eq: condnorm}) a small perturbative piece first obtained by Becchi et al
\cite{PSEUDO} (see also
\cite{SNB,BROAD,JAMIN}):
\beq
\la\bar ss\ra_{\msb}=\la\bar ss\ra-\frac{3}{2\pi^2}\frac{2}{7}\ga 
\frac{1}{a_s}-\frac{53}{24}\dr
\overline{m}_s^3.
\eeq
This leads to the ratio of the {\it non-normal ordered} condensates:
\beq
{\la \bar ss \ra}/{\la \bar uu \ra}\vert_{\msb} = 0.75\pm 0.12.
\eeq
The previous estimates are in good agreement with the ones from
chiral perturbation theory \cite{14} (see also \cite{JAMIN3}). They 
are also in fair agreement
with the one from the baryonic sum rules
\cite{DOSCH}--\cite{IOFFEBAR}, though we expect that the result from the latter
is less accurate due to the complexity of the analysis in this channel
(choice of the interpolating operators, eventual large effects of the continuum
due to the nearby Roper resonances,...).
%%%%%%%%%%%%%%%%%%%%%%%%%%%%%%%%%%%%%%%%%%%%%%%%%%%%
\subsection{$SU(2)$ breaking of the quark condensate}
The $SU(2)$ breaking of the quark condensate has been studied
for the first time in \cite{BRAMON}
and in \cite{PSEU,SNB}. Using similar approaches, the estimate is 
\cite{SNB,SNP}:
\beq
\la \bar dd\ra/\la\bar uu\ra\simeq 1-9\times 10^{-3}~.
\eeq
The previous estimate is in good agreement with the one from FESR \cite{PSEU}.
%%%%%%%%%%%%%%%%%%%%%%%%%%%%%%%%%%%%%%%%%%%%
\section{Heavy quark masses}
In the previous part of this book, we have already discussed the
different definitions of the
heavy quark masses and given their values. Contrary to the light
quark masses, the definition
of pole quark masses $p^2=M^2_H$ can (in principle) be introduced
perturbatively for heavy
quarks
\cite{COQUE,TARRA1,SN1} similarly to the one of the electron as here
the quark mass is much
heavier than the QCD scale $\Lambda$ such the perturbative approach
makes sense. However, a
complication arises due to the resummation of the QCD series
\cite{BENEKE} such that the pole
mass definition has an intrinsic ambiguity, which can be an obstacle
for its improved
accurate determination though the effect is relatively small.
Alternative definitions free from such
ambiguities have been proposed in the literature
\cite{POLE3,HOANG}. In this section, we shall discuss the determinations of the
perturbative running
quark masses which do not have such problems.
%%%%%%%%%%%%%%%%%%%%%%%%%%%%%%%%%%
\subsection{The quarkonia channel}
Charmonium and bottomium are the standard  channels for extracting
the charm and
bottom quark masses. Most of the sum rule analysis are based on the
$Q^2=0$ moments (MOM)
originally introduced by SVZ for the study of the charmonium systems:
\beq
{\cal M}_n\equiv \frac{1}{n!}\ga -{d\over dQ^2}\dr^n
\Pi\Big{|}_{Q^2=0}=\int_{4m^2}^\infty{dt\over t^{n+1}}{1\over\pi}{\rm
Im}\Pi(t)~,
\eeq
but convenient for the bottomium systems due to a much better
convergence of the OPE. In
\cite{SRRY}, the $Q^2\not= 0$ moments have been introduced for
improving the convergence
of the QCD series:
\beq
{\cal M}_n(Q^2_0)\equiv \frac{1}{n!}\ga -{d\over dQ^2}\dr^n
\Pi\Big{|}_{Q^2=Q^2_0}=\int_{4m^2}^\infty{dt\over
(t+Q^2_0)^{n+1}}{1\over\pi}{\rm
Im}\Pi(t)~,
\eeq
The spectral function can be related to the $e^+e^-\rar Q\bar Q$
total cross-section via
the optical theorem:
\beq
{\rm Im}\Pi(t+i\epsilon)=\frac{1}{12\pi Q^2_Q}{\sigma (e^+e^-\rar Q\bar Q)\over
\sigma (e^+e^-\rar \mu^+\mu^-)}~.
\eeq
$Q_Q$ is the heavy quark charge in units of e.
The contribution to the spectral function is as usual saturated by
the lowest few
   resonances plus the QCD continuum above the threshold $t_c$:
\beq
\mbox{Im}  \Pi_Q(t) =\frac{3}{4\alpha^2}\frac{1}{Q^2_Q}
\sum_{i}
{\Gamma_i M_i} \delta (t-M^2_i)~~+~~
\theta (t-t_c) \mbox{Im} \Pi^{QCD}_Q(t),
\eeq
where
$\Gamma_i$ is the electronic width of the resonances with the value
given in PDG \cite{PDG}. Retaining the observed resonances, the value of
$\sqrt{t_c}$ fixed from stability analysis is about $(11\sim12)$~GeV for the
$\Upsilon$-- and about 5 GeV for the $J\Psi$--families. However, the
result will be
practically independent from this choice of $t_c$ due to the almost
complete dominance of
the lowest ground state to the spectral function at the stability
point. An alternative
approach used in
\cite{SN1,SN2} is the LSR:
\beq
{\cal L}(\tau)=\int_{4m^2}^\infty dt~\exp^{-t\tau}{1\over\pi}{\rm Im}\Pi(t)~.
\eeq
This sum rule is particularly convenient for the analysis of the charmonium
systems as the corresponding OPE converges faster than the moment sum
rules. It has been
noticed in \cite{SN2} that the ratios of sum rules (and their finite
energy sum rule
(FESR) variants) are more appropriate for the estimate of the
quark mass as these ratios equate {\it directly} the mass squared of
ground state to that of the quark:
\beq
{\cal R}_n \equiv \frac{{\cal M}^{(n)}}{{\cal M}^{(n+1)}}~~~~~~~
\mbox{and}~~~~~~~
{\cal R}_\tau \equiv -\frac{d}{d\tau} \log {{\cal L}},
\eeq
   They also eliminate, to leading order,
some artefact dependence due to the sum rules (exponential weight factor
or number of derivatives) and some other systematic errors appearing
in each individual
moments. For the perturbative part,
we shall use (without expanding in 1/M)
the Schwinger extrapolation formula to two-loops:
\beq
\mbox{Im} \Pi^{pert}_Q(t)
\simeq \frac{3}{12\pi}v_Q\ga \frac{3-v^2_Q}{2} \dr
\aga 1+\frac{4}{3}\als f(v_Q) \adr ,
\eeq
where:
\beq
v_Q= \sqrt{1-4M^2_Q/t}
~,~~~~~~~~~~~~~~~~~
f(v_Q)=\frac{\pi}{2v_Q}-\frac{(3+v_Q)}{4}\ga \frac{\pi}{2}-\frac{3}{4\pi}
\dr
\eeq
are respectively the quark velocity and the Schwinger function
\cite{117}. We express this
spectral function in terms of the running mass by using the
two-loops relation given in  previous chapter and including the
$\als\log(t/M^2_Q)$-term appearing for off-shell quark.
We shall add to this perturbative
expression the lowest dimension $\la \als G^2 \ra $
non-perturbative effect (it is known as explained in the previous 
part of the book
%\ref{part: power corrections}
that, for a heavy-heavy quark correlator, the heavy quark
condensate contribution
is already absorbed into the gluon
one)
which among the available higher dimension
condensate-terms can only give a non-negligible contribution.
The gluon condensate contribution to the moments ${\cal M}^{(n)}$ and
so to ${\cal R}_n$ can be copied from the original work of SVZ \cite{SVZ}
and reads:
\beq
{\cal M}^{(n)}_G=-{\cal M}^{(n)}_{pert} ~\frac{(n+3)!}{(n-1)!(2n+5)}
\frac {4\pi}{9}\frac{\la \als G^2 \ra}{\ga 4M_Q^2 \dr^2},
\eeq
where ${\cal M}^{(n)}_{pert}$ is the lowest perturbative expression
of the moments.
The one to the Laplace ratio ${\cal R}_\tau $ can be also
copied from the
original work of Bertlmann \cite{BERTM}, which has been expanded
recently in $1/M_Q$ by \cite{DOMPAV}. It reads:
\beq
{\cal R}^G_\tau \simeq (4M^2_Q)\frac{2\pi}{3}\la \als G^2 \ra \tau^2
\ga 1+\frac{4}{3\omega}-\frac{5}{12\omega^2} \dr ,
\eeq
where $\omega \equiv 4M^2_Q \tau$. The results of the analysis from the
ratios of moments and Laplace sum rules give the values of the running masses
to order $\alpha_s$ \footnote{The inclusion of the
$\alpha_s^2$ correction is under study.} :
\beq
\bm_c(\bm_c)=(1.23\pm 0.03\pm 0.03)~{\rm GeV}~,~~~~~~~~~~~~~
\bm_b(\bm_b)=(4.23\pm 0.04\pm 0.02)~{\rm GeV}~,
\eeq
where the errors are respectively due to $\alpha_s(M_Z)=0.118\pm
0.006$ and $\la \als
G^2\ra=(0.06\pm 0.03)$ GeV$^4$ used in the original work. These
running masses can be
converted into the pole masses at this order. Non-relativistic
versions of these sum rules
(NRSR) introduced by \cite{VOLOSHIN} have also been used in
\cite{SN1,SN2} for determining the $b$
quark mass. These NRSR approaches have been improved by the inclusion
of higher order QCD corrections
and resummation of the Coulomb corrections from ladder gluonic 
exchanges. Some recent
different determinations  are given in Tables \ref{tab: mc} and \ref{tab: mb}.
%%%%%%%%%%%%%%%%%%%%%%%%%%%%%%%%%%%%%
\begin{table}[H]
\begin{center}
% space before first and after last column: 1.5pc
% space between columns: 3.0pc (twice the above)
\setlength{\tabcolsep}{0.2pc}
% -----------------------------------------------------
% adapted from TeX book, p. 241
%\newlength{\digitwidth} \settowidth{\digitwidth}{\rm 0}
%\catcode`?=\active \def?{\kern\digitwidth}
% -----------------------------------------------------
\caption{QSSR {\it direct determinations} of $\bm_c(\bm_c)$ in
$\msb$-scheme and of the pole
mass $M_c$ from $J/\Psi$-family, $e^+e-$ data and $D$-meson and
comparisons with lattice results.
Determinations from some other sources are quoted in PDG \cite{PDG}. The
results are given in units of GeV. The estimated error in the SR
average comes from an arithmetic average
of the different errors. The average for the pole masses is given at
NLO. The one
of the running masses is almost unchanged from NLO to NNLO
determinations. $\Longleftarrow$ means that
perturbative relation between the different mass definitions have
been used to get the quoted values.}
\label{tab: mc}
%\begin{tabular*}{\textwidth}{@{}l@{\extracolsep{\fill}}lcl}
\begin{tabular}{l l l l l}
%\hline
\hline
                &&& \\
   Sources&$\overline{m}_c (\bm_c)$&$M_c$&Comments& Authors\\
&&&\\
\hline
&&&\\
\boldmath $J/\Psi$\bf -family&&&&\\
%&&&&\\
MOM and LSR at NLO&$(1.27\pm 0.02)\Longleftarrow $&$(1.45\pm
0.05)$&$\Longleftarrow
m(-m^2_c)=(1.26\pm 0.02)$&SN89
\cite{SN1}\\ Ratio of LSR at NLO&$(1.23\pm 0.04)\Lrar $&$(1.42\pm
0.03)$&&SN94 \cite{SN2}\\
NRSR at NLO&$(1.23\pm 0.04)\Longleftarrow $&$(1.45\pm 0.04)$&&SN94 \cite{SN2}\\
SR at NLO&$(1.22\pm 0.06)\Longleftarrow $&$(1.46\pm 0.04)$&&DGP94
\cite{DOMPAV}\\
NRSR at NNLO&$(1.23\pm 0.09)$&$(1.70\pm 0.13)^* $&&EJ01 \cite{EIDEMUL}\\
&&&\\
\boldmath $e^+e^-$\bf data&&&&\\
%&&&&\\
FESR at NLO&$(1.37\pm 0.09)$&&&PS01 \cite{PS01}\\
MOM at NNLO&$(1.30\pm 0.03)$&&&KS01 \cite{KS01}\\
NLO&$(1.04\pm 0.04)\Longleftarrow$&$1.33\sim 1.4$&& M01 \cite{M01}\\
&\\
\boldmath $D$ \bf meson&\\
%&\\
Ratio of LSR at NNLO&$(1.1\pm 0.04) $&$(1.47\pm 0.04)$&&SN01 \cite{SNBC3}\\
&&&\\
\hline
&&&\\
{\bf SR Average}& $(1.23\pm 0.05)$&$(1.43\pm 0.04)$
\\ &&&\\
\hline
&\\
   \bf Quenched Lattice&&&&\\
%&&&&\\
%&$(1.53\pm 0.13)$&&&APE98 \cite{ape98}\\
&$(1.33\pm 0.08)$&&&FNAL98 \cite{fnal98}\\
&$(1.20\pm 0.23)$&&&NRQCD99 \cite{nrqcd99}\\
&$(1.26\pm 0.13)$&&&APE01 \cite{apemc}\\
%&\\
%\hline
&&&\\
\hline
\end{tabular}
\end{center}
{* Not included in the average.}
\end{table}
\nin
%%%%%%%%%%%%%%%%%%%%%%%%%%%%%%%%%%%%%%%%%%%%%%%%
%%%%%%%%%%%%%%%%%%%%%%%%%%%%%%%%%%%%%%%%%%%%%%%%%%%%%%
\subsection{The heavy-light $D$ and $B$ meson channels}
\begin{figure}[hbt]
\begin{center}
\includegraphics[width=9cm]{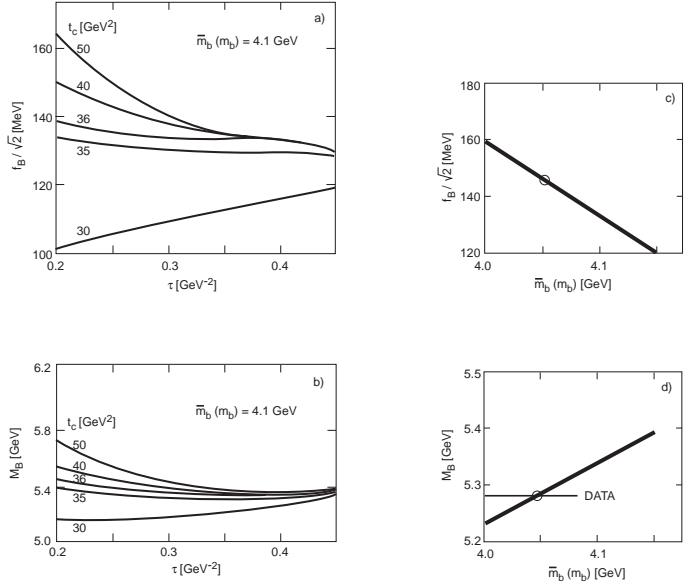}
\caption{Laplace sum rule analysis of $f_D$ and $\bm_c(\bm_c)$.}
\label{fig: mcfd}
\end{center}
\end{figure}

Heavy quark masses can also be extracted from the heavy-light quark
channels because the corresponding
correlators are sensitive to leading order to the values of these masses
\cite{SNB,SNMIXED,SN2,SNBC2,SNBC3}. Again, we shall be concerned
here with the LSR ${\cal L}(\tau)$ and
the ratio ${\cal R}(\tau)$. The latter sum  rule,
   or its slight modification, is useful, as it is equal to the
resonance mass squared, in
   the simple duality ansatz parametrization of the spectral function:
\beq
\frac{1}{\pi}\mbox{ Im}\psi_5(t)\simeq f^2_DM_D^4\delta(t-M^2_D)
   \ + \
   ``\mbox{QCD continuum}" \theta (t-t_c),
\eeq
where $f_D$ is the decay constant analogue to \footnote{Notice that
we have adopted here the lattice
normalization for avoiding confusion. We shall discuss its
determination in the next chapter.}
$f_\pi=130.56$ MeV. The QCD side of the sum rule reads:
\bea\label{eq: lsrlh}
{\cal L}_{QCD}(\tau)
&=& M^2_Q\Bigg{\{}\int_{M^2_Q}^{\infty}
{dt}~\mbox{e}^{-t\tau}~\frac{1}{8\pi^2}\Bigg{[} 3 t(1-x)^2\ga
1+\frac{4}{3}\as f(x)\dr+\as^2 R{2s}\Bigg{]}\nnb\\
&&~\,\, +\Big{[} C_4\la O_4\ra +C_6\la
O_6\ra\tau\Big{]}~\mbox{e}^{-M^2_Q\tau}\Bigg{\}}~,
\eea
where $R{2s}$ is the new $\alpha_s^2$-term obtained semi-analytically
in \cite{CHETP2} and is
available as  a Mathematica package program Rvs.m. Neglecting $m_d$,
the other terms are:
\bea
x&\equiv& M^2_Q/t,\nnb\\
f(x)&=&\frac{9}{4}+2\rm{Li}_2(x)+\log x \log (1-x)-\frac{3}{2}\log
(1/x-1)\nnb\\
& & -\log (1-x)+ x\log (1/x-1)-(x/(1-x))\log x, \nnb\\
C_4\la O_4\ra&=&-M_Q\la \bar dd\ra +\la \als G^2\ra/12\pi\nnb\\
C_6\la O_6\ra&=&\frac{M^3_Q\tau}{2}\ga 1-\frac{1}{2}M^2_Q\tau\dr
g\la\bar d\sigma_{\mu\nu}\frac{\lambda_a}{2}G_a^{\mu\nu}d\ra\nnb
\\ &&-\ga\frac{8\pi}{27}\dr\ga
2-\frac{M^2_Q\tau}{2}-\frac{M^4_Q\tau^2}{6}\dr\rho\als \la \bar
\psi\psi\ra^2~.
\eea
The previous sum rules can be expressed in terms of the running mass
$\bar{m}_Q(\nu)$
\footnote{It is clear that, for the non-perturbative terms which are
known to leading order
of perturbation theory, one can use either the running or the pole
mass. However, we shall see
that this distinction does not affect notably the present result.}.
  From this expression, one can
easily deduce the expression of the ratio ${\cal R}(\tau)$, where the
unkown decay constant
disappears, and from which we obtain the running quark masses:
\beq\label{mcrun}
\bar m_c(m_c)=(1.10\pm 0.04)~{\rm GeV} ~.
\eeq
The analysis is shown in Fig. \ref{fig: mcfd}, where a simulataneous 
fit of the decay
constant from ${\cal L}$ and of $\bm_c(\bm_c)$ from ${\cal R}$ is 
shown \footnote{We shall discuss
the decay constant in the next section.}.
%%%%%%%%%%%%%%%%%%%
%%%%%%%%%%%%%%%%%%
\nin
Our
optimal results correspond to the case where both stability in $\tau$
and in $t_c$ are reached.
However, for a more conservative estimate of the errors we allow
deviations from the stability
points, and we take:
\beq
t_c\simeq (6\sim 9.5)~{\rm GeV}^2~,~~~~~~~~~~~~\tau\simeq (1.2\pm
0.2)~{\rm GeV}^{-2}~,
\eeq
and where the lowest value of $t_c$ corresponds to the beginning of
the $\tau$-stability region. Values
outside the above ranges are not consistent with the stability criteria.
One can inspect
that the dominant non-perturbative contribution  is due to the dimension-four
$M_c\la
\bar dd\ra$ light quark condensate, and test that the OPE is not
broken by high-dimension condensates
at the optimization scale. However, the perturbative radiative
corrections converge slowly, as the value of
$f_D$ increases by 12\% after the inclusion of the
$\alpha_s$ correction and the sum of the lowest order plus
$\alpha_s$-correction increases by 21 \% after the
inclusion of the
$\alpha_s^2$ term, indicating that the total amount of corrections of
21\% is still a reasonnable
correction despite the slow convergence of the perturbative series,
which might be improved using a
resummed series. However, as the radiative corrections are both
positive, we expect that this slow
convergence  will not affect in a sensible way the final estimate. A
similar analysis
is done for the pole mass. The discussions presented previously apply also
here, including the one of the radiative corrections. We quote the
final result:
\beq\label{mcpole}
   M_c=(1.46\pm 0.04)~{\rm GeV} ~,
\eeq
where the error is slightly smaller here due to the absence of the
subtraction scale uncertainties.
One can cross-check that the two values of $\bar m_c(m_c)$ and $M_c$
give the ratio:
\beq
{M_c}/{\bar m_c(m_c)}\simeq  1.33~,
\eeq
   which satisfies quite well the
three-loop perturbative relation $M_c/\bar m_c(m_c)=1.33$. This could
be a non-trivial result if one
has in mind that the quark pole mass definition can be affected by
non-perturbative corrections not
present in the standard SVZ-OPE. In particular, it may signal that
$1/q^2$ correction of the type discussed in \cite{ZAKA,CNZ,SNZAK}, if
present, will only affect weakly the standard
SVZ-phenomenology as observed explicitly in the light quark, gluonia
and hybrid channels \cite{CNZ}.
Using analogous analysis for the $B$ meson, we obtain at the
optimization scale $\tau=0.375$
GeV$^{-2}$ and $t_c=38$ GeV$^2$:
\beq\label{mbrun}
   \bar m_b(m_b)=(4.05\pm 0.06)~{\rm GeV} ~,
\eeq
while using the pole mass as a free parameter, we get:
\beq\label{mbpole}
   M_b=(4.69\pm 0.06)~{\rm GeV} ~,
\eeq
One can again cross-check that the two values of $\bar m_b(m_b)$ and
$M_b$ lead to
\beq
{M_b}/{\bar
m_b(m_b)}=1.16~,
\eeq
to be compared with 1.15 from the three-loop perturbative relation
between $M_b$ and $\bm_b$, and
might indirectly indicate the smallness of the $1/q^2$ correction if
any. One can immediately notice
the agreement of the results from quarkonia and heavy-light quark
channels. Comparisons with other
determinations are given in Tables \ref{tab: mc} and  \ref{tab: mb}.

%%%%%%%%%%%%%%%%%%%%%%%%%%%%%%%%%%%%%
\begin{table}[hbt]
\begin{center}
% space before first and after last column: 1.5pc
% space between columns: 3.0pc (twice the above)
\setlength{\tabcolsep}{0.2pc}
% -----------------------------------------------------
% adapted from TeX book, p. 241
%\newlength{\digitwidth} \settowidth{\digitwidth}{\rm 0}
%\catcode`?=\active \def?{\kern\digitwidth}
% -----------------------------------------------------
\caption{The same as in Table \ref{tab: mc} but for the $b$-quark. }
\label{tab: mb}
%\begin{tabular*}{\textwidth}{@{}l@{\extracolsep{\fill}}lcl}
\begin{tabular}{l l l l l}
%\hline
\hline
                &&& \\
   Sources&$\overline{m}_b (\bm_b)$&$M_b$&Comments& Authors\\
&&&\\
\hline
&&&\\
\boldmath $\Upsilon$\bf -family&&&&\\
%&&&&\\
MOM and LSR at NLO&$(4.24\pm 0.05)\Longleftarrow $&$(4.67\pm
0.10)$&$\Longleftarrow
m_b(-m^2_b)=(4.23\pm 0.05)$&SN89
\cite{SN1}\\ Ratio of LSR at NLO&$(4.23\pm 0.04)\Lrar $&$(4.62\pm
0.02)$&&SN94 \cite{SN2}\\
NRSR at NLO&$(4.29\pm 0.04)\Longleftarrow $&$(4.69\pm 0.03)$&&SN94 \cite{SN2}\\
FESR at NLO&$(4.22\pm 0.05)\Longleftarrow $&$(4.67\pm 0.05)$&&SN95 \cite{SN2}\\
&$(4.14\pm 0.04)\Longleftarrow $&$(4.75\pm 0.04)$&&KPP98 \cite{KPP98}\\
NRSR at NNLO&$(4.20\pm 0.10)$&&&PP99, MY99 \cite{PEP99}\\
MOM at NNLO&$(4.19\pm 0.06)$&&&JP99 \cite{JP99}\\
NR at NNNLO&$(4.45\pm 0.04)$&&&PY00, LS00 \cite{PY00}\\
NR at NNNLO&$(4.21\pm 0.09)$&&&P01 \cite{P01}\\
NR at NNLO&$(4.25\pm 0.08)$&&$\Longleftarrow$ Residual mass&BS99 \cite{B99}\\
NR at NNLO&$(4.20\pm 0.06)$&&$\Longleftarrow$ $1S$ mass&H00 \cite{H00}\\
MOM at NNNLO&$(4.21\pm 0.05)$&&&KS01 \cite{KS01}\\
&\\
\boldmath $B$ \bf and \boldmath $B^*$ mesons&\\
%&\\
Ratio of LSR at NLO&$(4.24\pm 0.07)\Longleftarrow $&$(4.63\pm
0.08)$&&SN94 \cite{SN2}\\
Ratio of LSR at NNLO&$(4.05\pm 0.06) $&$(4.69\pm 0.06)$&~~$B$-meson
only&SN01 \cite{SNBC3}\\
&&&\\
\hline
&&&\\
%\multicolumn{1}{l}
{\bf SR Average}& $(4.24\pm 0.06)
$&$(4.66\pm 0.06)$&$\Lrar\bm_b(M_Z)=(2.83\pm 0.04)$\\ &&&\\
\hline
&\\
   \bf Average LEP &&&&\\
%&&&&\\
3-jets at $M_Z$&$(4.23\pm 0.94) $&&$\Longleftarrow\bm_b(M_Z)=(2.82\pm
0.63)$&LEP
\cite{RSB97}\\ &&&\\
   \bf Unquenched Lattice&&&&\\
%&&&&\\
&$(4.23\pm 0.09)$&&&APE00 \cite{APE00}\\
&\\
\hline
\end{tabular}
\end{center}
%\thanks{* Not included in the average.}
\end{table}
\nin
%%%%%%%%%%%%%%%%%%%%%%%%%%%%%%%%%%%%
\subsection*{Summary for the heavy quark masses and consequences}
  From Tables \ref{tab: mc} and  \ref{tab: mb}, we conclude that the
running $c$ and $b$ quark masses
to order $\alpha_s^2$ from the different sum rules analysis are likely to be:
\beq
\bm_c(\bm_c)=(1.23\pm 0.05)~{\rm
GeV}~,~~~~~~~~~~~~~~\bm_b(\bm_b)=(4.24\pm 0.06)~{\rm GeV}~,
\eeq
where the estimated errors come from the arithmetical average of
different errors. We have not tried
to minimize the errors from weighted average as the correlations
between these different determinations
are not clear at all. However, as one can see in the tables, the
quoted errors are typical for each
individual determinations. These results are consistent with other
determinations given in \cite{PDG}
and in particular with LEP average from three-jet events and lattice
values reported in the tables.
Using the previous relation between the short distance perturbative
pole and running masses, one
obtains, to order
$\alpha_s$:
\beq
M^{PT2}_c=(1.41\pm 0.06)~{\rm GeV}~,~~~~~~~~~~~~~~M^{PT2}_b=(4.63\pm
0.07)~{\rm GeV}~,
\eeq
and to order $\alpha_s^2$:
\beq
M^{PT3}_c=(1.64\pm 0.07)~{\rm GeV}~,~~~~~~~~~~~~~~M^{PT3}_b=(4.88\pm
0.07)~{\rm GeV}~,
\eeq
which are consistent with the average values to order $\alpha_s$
quoted in the tables and
in \cite{SN2}. However, one should notice the large effects due to
radiative corrections which can
reflect the uncertainties in the pole mass definition.
   From the previous values of the running
masses, one can also deduce the values of the RG invariant masses to order
$\alpha_s^2$:
\beq
\hat m_c=(1.21\pm 0.07)~{\rm GeV}~,~~~~~~~~~~~~~~\hat m_b=(6.9\pm
0.2)~{\rm GeV}~.
\eeq
We have used $\Lambda_4=325\pm 40$ MeV and $\Lambda_5=225\pm 30$ MeV.
Taking into account threshold
effects and using matching conditions, we can also evaluate the
running masses at the scale 2 GeV
and obtains:
\beq
\bm_c(2)=(1.23\pm 0.05)~{\rm GeV}~,~~~~~~~~~~~~~~\bm_b(2)=(5.78\pm
0.26)~{\rm GeV}~,
\eeq
Combining the values of $m_b$ and $m_s$ obtained in the previous
section, one can deduce the
scale independent mass ratio:
\beq
{m_b\over m_s}= 48.8\pm 9.8~,
\eeq
which is useful for model buildings.\\ One can also run the value of
$m_b$ at the $Z$-mass, and obtains
the value of $
\bm_b(M_Z)$ quoted in the table:
\beq
\bm_b(M_Z)=(2.83\pm 0.04)~{\rm GeV}.
\eeq
This value compares quite well with the ones measured at $M_Z$ from
three-jet heavy quark production
at LEP where the average $(2.83\pm 0.04)$~GeV of different
measurements \cite{RSB97} is given also in the table.
This is a {\it first indication} for the running of $m_b$ in favour
of the QCD predictions based
on the renormalization group equation.
%%%%%%%%%%%%%%%%%%%%%%%%%%%%
%%%%%%%%%%%%%%%%%%%%%%%%%%%%%%%%%%%%%%%%%%%%%%%%%%%%%%%%%%%%%%%%%%
  \section{The weak leptonic decay constants\index{decay constant} 
$f_{D_{(s)}}$ and $f_{B_{(s)}}$}
\begin{figure}[hbt]
\begin{center}
\includegraphics[width=5cm]{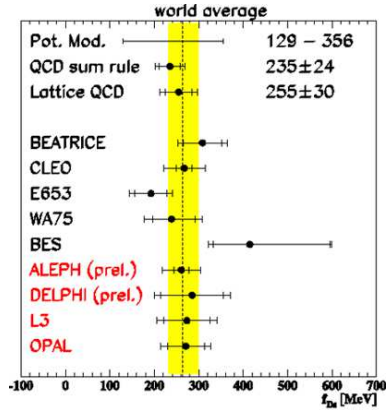}
\caption{Different measurements of $f_{D_s}$ compared with 
theoretical predictions from \cite{JEFF}.}
\label{fig: fds}
\end{center}
\end{figure}

In this section \footnote{This is an extension and an update of the 
some parts of the reviews given
in \cite{SNSRH}.}, we summarize the different results obtained from 
the QCD spectral
sum rules (QSSR) on the leptonic decay constants of the $B$ and $D$ 
mesons\index{meson} which are useful in
the analysis of the leptonic decay and on the $B$-$\bar B$ mixings. Intensive
activities have been devoted to this subject during the last few 
years using QSSR and lattice calculations.\\
The leptonic constant of the pseudoscalar\index{pseudoscalar} 
$P\equiv D,~B$ meson\index{meson} is defined
as:
\beq
\la 0|\partial_\mu A^\mu |P \ra = f_P M^2_P \vec{P}~,
\eeq
where $\vec{P}$ is the pseudoscalar\index{pseudoscalar} 
meson\index{meson} field and $f_P$ is the
pseudoscalar\index{pseudoscalar} decay constant\index{decay constant} 
which controls the $P \rar l\nu$ leptonic
decay width, normalized as $f_\pi=130.56$ MeV \footnote{In this 
chapter, we adopt this normalization used by
the lattice  and experimental groups. In the previous sections, we 
have used $f_\pi\equiv f_\pi/\sqrt{2}$ MeV.}.  The
current :
\beq
\partial_\mu A^\mu(x)^i_j=(m_i+M_j) \bar{\psi}_i (i\gamma_5)\psi_j~ ~
(i\equiv u,~d,~s;~j\equiv c,b)~,
\eeq
is the divergence\index{divergence} of the axial current. In the sum 
rule analysis, we
shall be concerned with the pseudoscalar\index{pseudoscalar} 
two-point correlator:
\beq
\Psi_5(q^2)= i\int d^4 x e^{iqx} \la 0|{\bf T} \partial_\mu
A^\mu (x)^i_j
  \ga\partial_\mu A^\mu (0)^i_j\dr^\dagger |0\ra ~.
\eeq
In the case of the $B(\bar ub)$
meson\index{meson}, the decay width into $\tau\nu_\tau$ reads:
\beq
\Gamma(B\rar\tau\nu_\tau~+~B\rar\tau\nu_\tau\gamma)=\frac{G^2_F|V_{ub}|^2}{4\pi}M_B\ga
1-\frac{M^2_\tau}{M^2_B}\dr^2M^2_\tau f^2_B~,
\eeq
where $M_\tau$ expresses the helicity suppression of the decay rate 
into light leptons
$e$ and $\mu$. This expression  shows that a good determination of 
$f_B$ will allow a
precise extraction of the CKM mixing angle\index{mixing angle} 
$V_{ub}$. One the other
hand,
$f_B$ and the so-called bag parameter $B_B$ also control the matrix
element of the $\Delta B=2$
$B^0$-$\bar{B}^0$ mixing matrix, which is of a non-perturbative
origin, as we shall discuss in another chapter.\\
However, contrary to the case of the $\pi$ and $K$ 
mesons\index{meson}, the leptonic
width of the heavy meson\index{heavy meson}\index{meson} is small as 
the corresponding
decay constant\index{decay constant}
vanishes as $1/\sqrt{M_Q}$, while the presence  of the neutrino in the
final state renders difficult the reconstruction of the signal and the
rejection of background. Moreover, the $B$ leptonic rate is Cabibbo suppressed,
which makes it unreachable with present measurements.
($\sim |V_{ub}|^2$), while the $D_s$ leptonic rate is Cabbibbo favoured
($\sim |V_{cs}|^2$). Recent measurements of $f_{D_s}$ are given in 
Fig. \ref{fig: fds}, where
the quoted average is \cite{JEFF}:
\beq
f_{D_s} \simeq (264\pm 37)~{\rm MeV}~.
\eeq
%%%%%%%%%%%%%%%%%%%
%%%%%%%%%%%%%%%%%%
%%%%%%%%%%%%%%%%%%%%%%%%%%%%%%%%%%%%%%%%%%%%%
\subsection{Upper bound on the value of $f_D$}
Within the QSSR framework, the decay constants\index{decay constant} of the $B$
and $D$ mesons\index{meson}
have been firstly estimated in \cite{102}, while the first upper bounds on
their values have been derived in \cite{WSRQCD} and updated
in the recent review \cite{SNSRH}.
Indeed, a {\it rigorous}
upper bound on these couplings\index{hadrons couplings} can be derived from the
second-lowest superconvergent\index{superconvergent sum 
rules}\index{convergent} moment:
\beq
{\cal M}^{(2)} \equiv \frac{1}{2!}\frac{\partial^2 \Psi_5(q^2)}
{\ga \partial q^2\dr^2} \Bigg{\vert} _{q^2=0}~,
\eeq
where for this low-moment, the OPE\index{Operator Product Expansion 
(OPE) } well behaves.
Using the positivity of the higher-state contributions to the
spectral function, one can deduce \cite{WSRQCD,BROAD}:
\beq
f_P \leq \frac{M_P}{4\pi} \aga 1+ 3 \frac{m_q}{M_Q}+
0.751 \bar{\alpha}_s+... \adr~,
\eeq
where one should not misinterpret the mass-dependence in this
expression compared to the one expected from heavy-quark\index{heavy 
quark} symmetry.
Applying this result to the $D$ meson\index{meson}, one obtains:
\beq
f_D \leq 2.14 f_\pi~,
\eeq
which is not dependent to leading order on the value of the 
charm\index{charm} quarkm mass.
Although
presumably quite weak, this bound, when combined with the recent
determination of the $SU(3)_F$ breaking\index{$SU(3)_F$ breaking} 
effects to two loops on
the ratio of decay constants\index{decay constant} \cite{SNFD}:
\beq
\frac{f_{D_s}}{f_D} \simeq (1.15 \pm 0.04)f_\pi~ ,
\eeq
implies
\beq
f_{D_s} \leq (2.46 \pm 0.09)f_\pi\simeq (321.2\pm 11.8)~{\rm MeV}~ ,
\eeq
which is useful for a comparison with the recent measurement of $f_{D_s}$,
with the average value given previously.
One cannot push, however, the uses
of the moments to higher $n$ values in this $D$ channel, in order to
minimize the continuum contribution to the sum rule with the aim to
derive an estimate of the decay constant\index{decay constant} from 
this method, and to derive
its ``correct" mass dependence, because the QCD series
will not converge at higher $n$ values.
%%%%%%%%%%%%%%%%%%%%%%%%%%%%%%%%%%%%%%%%%
\subsection{Estimate of the $D$ decay constant\index{decay constant} $f_D$}
The decay constant $f_D$ can be extracted from the pseudoscalar 
Laplace sum rules given in Eq. (\ref{eq:
lsrlh}) \footnote{For reviews, see e.g. \cite{DOMIREV}.}. Prior 1987, 
the different sum rules estimate of the
decay constant\index{decay constant}
$f_P$ have been inconsistent among each others. To our knowledge,
the first attempt to understand such disrepancies has been done in
\cite{SNFBI,SNF} (see also \cite{DOMINGUEZ}), where it has been 
shown, {\it for the
first time} and a long time before the lattice results, that:
\beq
f_D\approx f_B \approx (1.2\sim 1.5)f_\pi~,
\eeq
which differs from that expected from the heavy quark symmetry scaling
law \cite{HQSYM}:
\beq
f_B\approx \sqrt{\frac{M_D}{M_B}}f_D\ga
\frac{\alpha_s(M_c)}{\alpha_s(M_b)}\dr^{-1/{\beta_1}}~,
\eeq
valid in the extreme case where the heavy quark mass\index{heavy 
quark mass}\index{quark mass} is infinite
\footnote{Finite mass corrections to this formula will be dicussed later
on.}.
  It has also been understood that the apparent disagreement among different
existing QSSR numerical results in the literature is
{\it not only} due to the choice of the continuum threshold 
$t_c$\index{threshold} [ its effect is $(7\sim
10)\%$ of the result when one moves $t_c$ from the one at the 
beginning of sum rule variable to the one
where  the $t_c$ stability is reached.]\footnote{In some papers in 
the literature, the value of $t_c$ is
taken smaller than the previous range. In this case, the $t_c$ effect is
larger than the one given here.} as
misleadingly claimed in the literature. Indeed, the main effect is
{\it also} due to the different values of the quark mass\index{quark 
mass}es used
\footnote{A critical review on the
discrepancy between different existing estimates is given in
\cite{SNREV}.}, which is shown explicitly
in the table of \cite{SNFD}.
\\  In the
$D$ channel, the most appropriate sum rule is the (relativisitc) Laplace
sum rule\index{Laplace sum rule}, as the OPE\index{Operator Product 
Expansion (OPE) } of the
$q^2=0$ moments does not converge for larger value of the number of
derivatives $n$, at which the $D$ meson\index{meson} contribution to 
the spectral integral
is optimized.  The results from different groups are consistent with each
others for a given value of the
$c$-quark mass\index{quark mass}. For the $D$ meson\index{meson}, the 
optimal result is obtained for:
\beq
6\leq t_c \leq 9.5~\mbox{GeV}^2~,~~~~~~~~~\tau\simeq (1.2\pm 
0.2)~\mbox{GeV}^{-2}~.
\eeq
where the QCD corrections are still reasonnably small. The most 
recent estimate including $\alpha_s^2$
corrections from a simultaneous fit of the set either 
($f_D,\bm_c(m_c)$) or ($f_D, M_c^{pole}$) is given in
Fig. \ref{fig: mcfd}. The obtained values of the quark masses have 
been quoted in Table \ref{tab:
mc}. The resulting value of $f_D$ is \cite{SNBC3}:
\beq
f_D \simeq (203\pm 23)~{\rm MeV}~,
\eeq
in agreement with the recent evaluation $(195\pm 20)$ MeV at order 
$\alpha_s^2$ but using the
pole mass as input \cite{STEIN01}.
%%%%%%%%%%%%%%%%%%%%%%%%%%%%%%%%%%%%%%%%%%%%%%%%%%%%%%%%%%%%%%%%%%%%%%%%
\subsection{Ratio of the decay constants $f_{D_s}/f_D$ and $f_{B_s}/f_B$}
The $SU(3)$ breaking ratios
$f_{D_s}/f_D$ and $f_{B_s}/f_B$ have been obtained semi-analytically in
\cite{SNFD}. In order to have a qualitative understanding of the size 
of these corrections, we strat from
the global hadron-quark duality sum rule:
\beq
\int_0^{\omega_c}d\omega~ {\rm Im}\Psi_5^{res}(\omega)\simeq
\int_0^{\omega_c}d\omega ~{\rm Im}\Psi_5^{\bar qQ}(\omega)~,
\eeq
where $\omega_c$ is the continuum energy defined as:
\beq
t=(E+M_Q)^2\equiv M^2_Q+\omega M_Q~.
\eeq
Keeping the leading order terms in $\alpha_s$ and in $1/M_Q$, it leads to:
\beq
R_P\simeq \rho_P\Bigg{\{}1+3\ga{m_s\over \omega_c}\dr\ga 1-{m_s\over M_Q}\dr
-6\ga{m_s\over \omega_c}^2\dr-\ga{m_s\over M_Q}\dr \ga 1-{m_s\over 
M_Q}\dr\Bigg{\}}~,
\eeq
where:
\beq
\rho_P\equiv \ga{M_P\over M_{P_s}}\dr^2\ga 1+{m_s\over M_Q}\dr~.
\eeq
The value of $\omega$ is fixed from stability criteria to be 
\cite{SNFBI,SNREV,SNHQET2,SNHQET}:
\beq
\omega_c\simeq (3.1\pm 0.1)~{\rm GeV}.
\eeq
The sum rule indicates that the $SU(3)$ breaking corrections are of 
two types, the one $m_s/M_Q$ and the
other $m_s/\omega_c$. More quantitatively, we work with the Laplace sum rule:
\beq
{\cal L}=\int_0^{\omega_c}d\omega~{\rm e}^{- \omega\tau}~{\rm 
Im}\Psi_5^{res}(\omega)
\eeq
Analogously the Laplace sum rule gives:
\beq
R_P^2\simeq \rho_P^2\Bigg{\{}1+2(2.2\pm 0.2)\ga{m_s\over 
\omega_c}\dr\ga 1-{m_s\over M_Q}\dr
-2\ga 8.2\pm 1.6\dr\ga{m_s\over \omega_c}^2\dr\Bigg{\}}~,
\eeq
where the numerical integration includes a slight $M_Q$ dependence.
Including $m_s\alpha_s$ and $m_s^2\alpha_s$-corrections, the 
resulting values of the ratio are:
\beq
R_D\equiv {f_{D_s}\over f_D}=1.15\pm 0.04~,~~~~~~~ R_B\equiv 
{f_{B_s}\over f_B}=1.16\pm 0.05~.
\eeq
This result implies:
\beq
f_{D_s} \simeq (235\pm 24)~{\rm MeV}~,
\eeq
which agrees within the errors with the data \cite{JEFF}
and lattice \cite{LATTFB} averages both quoted in Fig.~\ref{fig: fds}.
This feature increases the confidence in the uses of the QSSR method 
for predicting the not yet measured decay
constant\index{decay constant} of the $B$ meson\index{meson}.
%%%%%%%%%%%%%%%%%%%%%%%%%%%%%%%%%%%%%
\begin{table}[hbt]
\begin{center}
% space before first and after last column: 1.5pc
% space between columns: 3.0pc (twice the above)
\setlength{\tabcolsep}{0.4pc}
% -----------------------------------------------------
% adapted from TeX book, p. 241
%\newlength{\digitwidth} \settowidth{\digitwidth}{\rm 0}
%\catcode`?=\active \def?{\kern\digitwidth}
% -----------------------------------------------------
\caption{Estimate of $f_{B_{(s)}}$ to order $\alpha_s^2$ and 
$f_{B_{s}}/f_{B}$ to order
$\alpha_s$ from QSSR  and comparison  with the lattice.}
\label{tab: fb}
%\begin{tabular*}{\textwidth}{@{}l@{\extracolsep{\fill}}lcl}
\begin{tabular}{l l l l l l}
%\hline
\hline
               &&& \\
  Sources&$f_{B}$ [MeV]&$f_{B_{(s)}}/f_{B}$&$f_{B_s}$ [MeV]&Comments & Authors\\
&&&\\
\hline
&&&\\
\bf QSSR&&&&\\
LSR&$203\pm 23$& $1.16\pm 0.04\Lrar$&$236\pm 30$&$M_{pole}$, $\bm_b$: 
output &SN01 \cite{SNBC3} \\
&$210\pm
19$& &$244\pm 21$&$\bm_b$: input&JL01 \cite{JL01} \\
HQETSR&$206\pm
20$& &&$M_{pole}$: input&PS01 \cite{STEIN01} \\
%&&&&\\
&&&\\
\hline
&&&\\
{\bf SR Average}& $207\pm 21\Lrar$&&$ 240\pm 24$&
\\ &&&\\
\hline
&&&\\
\bf Unq. Lattice&&&&\\
&$200\pm 30$& $1.16\pm 0.04\Lrar$&$232\pm 35$&average &LAT01\cite{LAT01}
\\
&\\
\hline
\end{tabular}
\end{center}
%\thanks{* Not included in the average.}
\end{table}
\nin
%%%%%%%%%%%%%%%%%%%%%%%%%%%%%%%%%%%%%%%%%%%%%%%%%%%%%%%%%%%%%%%%%%%%%%%%%%%
\subsection{Estimate of the $B$ decay constant\index{decay constant} $f_B$}
For the estimate of $f_B$, one can either work with the Laplace, the
moments or
their non-relativistic variants because the $b$-quark 
mass\index{quark mass} is relatively
heavy. The optimal result
which we shall give here comes from the Laplace relativistic sum
rules. They corresponds to the {\it conservative range} of parameters:
\beq
  0.6\leq E_c^{(b)}\equiv \sqrt{t_c}-M_B\leq 1.8~\mbox{GeV}~,
~~~~~~~~~\tau\simeq 0.38~\mbox{GeV}^{-2}~,
\eeq
which have been used in the previous section for getting the $b$-quark mass.
As shown in the figure given in \cite{SNF1,SNF2}, the
dominant corrections come from the $\la \bar uu\ra$ quark 
condensate\index{quark condensate}
with the strengh $(30\sim
40)\%$ of the lowest order term in $f_B$, while the higher condensate 
effects are
smaller, which are respectively $-(20\sim 30)\%~,~ +(5\sim 8)\%$ for the
$d=5$ and 6 condensates. This shows, despite the large value of the quark
condensate\index{quark condensate}
contribution, that the OPE\index{Operator Product Expansion (OPE) } is
convergent\index{convergent}. It has been noticed in \cite{SNF1,SNF2},
that the convergence of the OPE\index{Operator Product Expansion (OPE) }
is faster for the relativistic LSR\index{Laplace sum rule}  than for the
moments, such that the most precise result should from the 
LSR\index{Laplace sum rule} .
In both cases the perturbative corrections are small.
One obtains from the relativistic LSR\index{Laplace sum rule} , the 
results to order $\alpha_s^2$
\cite{SNBC3}:
\beq
f_B \simeq (203\pm 23)~{\rm MeV}\simeq (1.55\pm 0.18)f_\pi ~,
\eeq
and to order $\alpha_s$ (see previous discussion) \cite{SNFD}:
\beq
  \frac{f_{B_s }}{f_B}\simeq 1.16 \pm 0.04~.
\eeq
These values of $f_B$ and $f_D$ agree quite well with the previous QSSR
findings in \cite{SNFD}, \cite{SNB} and \cite{DOMINGUEZ}. They also 
agree with the
non-relativistic sum rules estimate in the full theory \cite{SNFBI}, in HQET
\cite{GROZINB,STEIN01} and  in \cite{SNHQET2,SNHQET}. However, unlike
the relativistic sum rules, the HQET sum rule is strongly affected by the huge
perturbative radiative corrections of about 100$\%$, which is 
important at the sum
rule scale of about 1 GeV at which the HQET sum rule optimizes. These 
results are also
in good agreement with the lattice average estimate given in Table 
\ref{tab: fb}.
%%%%%%%%%%%%%%%%%%
\subsection{Static limit and $1/M_b$-corrections to $f_B$}
As noticed previously, the {\it first} result
$f_B \simeq f_D$ in \cite{SNFD}, which has been confirmed by recent estimates
from different approaches, shows a large violation of the scaling law
expected from heavy-quark\index{heavy quark} symmetry. This result 
suggests that finite
quark mass\index{quark mass} corrections are still huge at the $D$ and $B$
meson\index{meson} masses. The first attempt to understand this 
problem analytically is
in \cite{SNF} in terms of large corrections of the type $E_c/M_b$ if 
one expresses in this paper
the continuum threshold $t_c$ in terms of the threshold energy $E_c$:
\beq
t_c\equiv (E_c+M_b)^2.
\eeq
Later on different approaches have been investigated for the estimate 
of the size
of these corrections. \\
In the lattice approach, these mass corrections have been estimated from
a fit of the obtained value of the meson\index{meson} decay 
constant\index{decay constant}
at finite and
infinite (static limit\index{static limit}) values of the heavy quark
mass\index{heavy quark mass}\index{quark mass} and
by assuming that these corrections are polynomial in $1/M_Q$ up to 
log. corrections. A
similar analysis has been done with the sum rule in the full theory
\cite{SNF1,SNF2}, by studying numerically the quark mass\index{quark 
mass} dependence of
the decay constant\index{decay constant} until the quark mass\index{quark mass}
value ($M_Q\leq 15$ GeV) until
which one may expect that the sum rule analysis is still valid. In so doing, we
use the parametrization:
\beq
f_B \sqrt{M_B} \simeq \tilde{f}_B
\alpha_s^{1/{\beta_1}}\aga 1-\frac{2}{3}\frac{\alpha_s}{\pi}-
\frac{A}{M_b}+\frac{B}{M^2_b}\adr~,
\eeq
by including the quadratic mass corrections. The analysis gives
\footnote{The numbers given in \cite{SNF1,SNF2} correspond to the
quark mass\index{quark mass} $M_b=4.6$ GeV and should be rescaled 
until the meson\index{meson}
mass $M_B$. In the following, we shall also work with $f_B$ 
normalized to be $\sqrt{2}$ bigger than
in the original papers.}:
\beq
\tilde{f}_B\equiv \ga f_B\sqrt{M_B}\dr_\infty\simeq
(0.65 \pm 0.06)~\mbox{GeV}^{3/2}~,
\eeq
which one can compare with the results from the HQET Laplace sum 
rule\index{Laplace sum rule}
\cite{GROZINB} and \cite{SNHQET,ZAL2}:
\beq
\tilde{f}_B\simeq
(0.35 \pm 0.10)~\mbox{GeV}^{3/2}~,
\eeq
and from FESR\index{finite energy sum rule} \cite{SNHQET}:
\beq
\tilde{f}_B\simeq
(0.57\pm 0.10)~\mbox{GeV}^{3/2}~,
\eeq
Taking the average of these three (independent) determinations,
one can deduce:
\beq
\tilde{f}_B\simeq
(0.58\pm 0.09)~\mbox{GeV}^{3/2}~,
\eeq
where we have done an arithmetic average of the errors.
This result is in good agreement with the lattice value
given in \cite{LATTFB,COLLINS} using nonperturbative clover fermions. 
One can translate
this result into the value of $f_B$ in the static limit approximation:
\beq
f^{stat}_B\simeq (1.9\pm 0.3)f_\pi~.
\eeq
We can also use the previous value of $f^{stat}_B$ together with the previous
values of $f_B$ and $f_D$ at the ``physical" quark mass\index{quark 
mass}es in order to determine
numerically the coefficients $A$ and $B$ of the $1/M_b$ and $1/M^2_b$ 
corrections.
In so doing, we use the values of the quark "pole" mass\index{quark 
mass}es given in Tables
\ref{tab: mc} and \ref{tab: mb}.  Then, we obtain from a quadratic fit:
\beq
  A\approx 0.98~
\mbox{GeV}~~~~~\mbox{and}~~~~~ B \approx 0.35~\mbox{GeV}^2~,
\eeq
while a linear fit gives a large uncertainty:
\beq
A\approx (0.74\sim 0.91)~\mbox{GeV}~.
\eeq
One can notice that the fit of these coefficients depends strongly
one the input values of $f_D$ and $f_B$. Indeed, using some other set of values
as in \cite{SNF1,SNF2,SNHQET2}, one obtains values about 2 times 
smaller.Therefore,
we consider as a conservative value of these coefficients an 
uncertainty of about 50\%:
The  value of A obtained is comparable with the one from HQET sum rules
\cite{NEUBERT2} and \cite{SNF1,SNHQET2} of about $0.7\sim 1.2$ GeV.
A similar value of $A$ has been
also obtained from the  recent NR lattice calculations with dynamical 
fermions and using
a linear fit \cite{COLLINS}:
\beq
  A\approx 0.7~\mbox{GeV}~.
\eeq
One can $qualitatively$ compare this result with the one obtained from
the analytic expression of the moment in the full theory \cite{SNB},
\cite{SNF,ZAL2}:
\beq
f_B^2 \approx \frac{1}{\pi^2}\frac{E^{3}_c}{M_b}
\Bigg{[} 1-\frac{2}{3}\frac{\alpha_s}{\pi}-\frac{3(n+2)E_c}{M_b}+...
-\frac{\pi^2}{2} \frac{\la \bar uu \ra}{E^3_c}+...\Bigg{]}~ .
\eeq
Here, one can notice that the size of the $1/M_b$ corrections depends on the
number of moments, such that their estimate using literally the 
expression of the
moments can be inconclusive. A qualitative estimate of these corrections can be
done from the semilocal duality\index{local duality} sum rule, which has more
intuitive physical meaning due
to its direct connection with the leptonic width and total cross-section
through the optical theorem. It corresponds to $n=-2$, and leads to the
$interpolating~formula$
\cite{ZAL2}:
\beq
\sqrt{2}f_B \sqrt{M_B} \approx  \frac{E^{3/2}_c}{\pi} \alpha_s^{1/{\beta_1}}
\ga \frac{M_b}{M_B}\dr^{3/2} \Bigg\{ 1-\frac{2}{3}\frac{\alpha_s}{\pi}
+\frac{3}{88}\frac{E^2_c}{M^2_b}
-\frac{\pi^2}{2} \frac{\la \bar uu \ra}{E^3_c}+...\Bigg\}~,
\eeq
from which, one obtains:
\bea
A &\approx &\frac{3}{2}(M_B-M_b) \simeq 1~\mbox{GeV},\nnb \\
B &\approx& \frac{3}{88}E_c^2+\frac{27}{32}(M_B-M_b)^2 \simeq 0.45~\mbox
{GeV}^2~,
\eea
which is in good agreement with the previous numerical estimate.
%%%%%%%%%%%%%%%%%%%%%%%%%%%%%
\section{Conclusions}
We have reviewed the determinations of the quark masses and leptonic 
decay constants of (pseudo)scalar mesons
which are useful in particle physics phenomenology. The impressive 
agreements of the QSSR results with the data
when they  are measured and/or with lattice calculations in different 
channels indicate the self-consistency
of the QSSR approach making it as one of the most powerful 
semi-analytical QCD nonperturbative approach
available today. Applications of these results for studying the 
$B$-$\bar B$ mixings and $CP$-violation will be
discussed in the next chapter.
\section*{Acknowledgements}
It is a pleasure to thank Valya Zakharov for the hospitality at the Max-Planck
Institute f\"ur Physik (Munich) where this work has been completed 
and for some discussions.

\end{document}